\documentclass[twocolumn,a4paper]{autart}    % Enable this line and disable the 
\usepackage[english]{babel}															% English language/hyphenation
\usepackage{graphics} % for pdf, bitmapped graphics files
\usepackage{mathtools} % assumes new font selection scheme installed
\usepackage{times} % assumes new font selection scheme installed
\usepackage{amsmath} % assumes amsmath package installed
\usepackage{amssymb}  % assumes amsmath package installed
\usepackage{amsfonts}%
\usepackage{euscript}
\usepackage{mathrsfs}
\usepackage[inline]{enumitem}
\usepackage{xcolor}
\usepackage{multirow,tabularx}
\usepackage{subcaption}
\usepackage{hhline}
\usepackage{algcompatible}
\usepackage{algorithm}
\usepackage{cite}
\usepackage{eufrak}
\usepackage[normalem]{ulem} % necessary for striketrough
\PassOptionsToPackage{hyphens}{url}
\usepackage[
  bookmarks=false,
  pdfpagelabels=false,
  hyperfootnotes=false,
  hyperindex=false,
  pageanchor=false,
  implicit=false,
  colorlinks=false,
]{hyperref}
\usepackage{tikz}

\usetikzlibrary{shadows.blur,positioning,backgrounds,fit,spy} %external
\usepackage{pgfplots}
\pgfplotsset{compat=newest}
\pgfplotsset{plot coordinates/math parser=false}

\def\arXivPrint{1}  % Comment out to exclude the arXiv.org title page for IEEE

\usepackage{xifthen} % better if then commands for defining commands
\usepackage{euscript}

\DeclareMathAlphabet{\mathpzc}{OT1}{pzc}{m}{it}
\DeclareMathAlphabet{\mathbscr}{U}{BOONDOX-cal}{b}{n}
\DeclareMathAlphabet{\mathdutchcal}{U}{dutchcal}{m}{n}
\DeclareMathAlphabet{\mathbdutchcal}{U}{dutchcal}{b}{n}
\DeclareMathAlphabet{\eucalit}{U}{euf}{m}{n}

\definecolor{iavO}{RGB}{230,150, 44}		% orange (high voltage cable)
\definecolor{iavH}{RGB}{162,0,103}			% raspberry
\definecolor{iavB5}{RGB}{ 12, 56,104}		% blue 5 (dark) - basic color for faces
\definecolor{iavB1}{RGB}{ 93,174,219}		% blue 1 (light)

\DeclareMathOperator*{\diag}{diag}

\newcommand{\NX}{{n_\mathrm{x}}}
\newcommand{\NXp}[1][]{%
\ifthenelse{\isempty{#1}}{{n_\mathrm{x}}}{{n_\mathrm{x}^{#1}}}%
}

\newcommand{\NY}[1][]{%
\ifthenelse{\isempty{#1}}{{n_\mathrm{y}}}{{n_\mathrm{y}^{#1}}}%
}
\newcommand{\NU}{{n_\mathrm{u}}}
\newcommand{\NP}{{n_\mathrm{p}}}
\newcommand{\NZ}{{n_\mathrm{Z}}}

\newcommand{\NPSI}{{n_{\psi}}}			% Additional {} such that it can be used in super and subscript

\newcommand{\NA}{{n_\mathrm{a}}}
\newcommand{\NB}{{n_\mathrm{b}}}
\newcommand{\NC}{{n_\mathrm{c}}}

\newcommand{\NTH}{{n_\mathrm{\theta}}}

\newcommand{\NMU}{{n_\mathrm{\mu}}}

\newcommand{\NMUTp}[1][]{%
\ifthenelse{\isempty{#1}}{{n_{\tilde{\mathrm{\mu}}}}}{{n_{\tilde{\mathrm{\mu}}}^{#1}}}%
}

\newcommand{\sU}{\mathbb{U}}
\newcommand{\sP}{\mathbb{P}}
\newcommand{\sX}{\mathbb{X}}
\newcommand{\sY}{\mathbb{Y}}
\newcommand{\sZ}{\mathbb{Z}}

\newcommand{\sS}{\mathbb{S}}

\newcommand{\Ru}{\mathbb{R}^{\NU}}
\newcommand{\Rp}{\mathbb{R}^{\NP}}
\newcommand{\Rx}{\mathbb{R}^{\NX}}
\newcommand{\Ry}{\mathbb{R}^{\NY}}

\newcommand{\ind}[1]{{[#1]}}

\newcommand{\sys}{\mathdutchcal{S}}

\newcommand{\M}{\mathtt{R}}
\newcommand{\N}{\mathtt{O}}

\newcommand{\reach}{R}
\newcommand{\obsv}{O}

\newcommand{\reachTV}{\mathbdutchcal{R}}
\newcommand{\obsvTV}{\mathbdutchcal{O}}

\newcommand{\hank}{\mathcal{H}}

\newcommand{\expct}{\mathbb{E}}

\newcommand{\rank}{\mathrm{rank}}

\newcommand{\SNR}{\mathrm{SNR}}
\newcommand{\BFR}{\mathrm{BFR}}
\newcommand{\Tr}{{\mathrm{Tr}\,}}
\newcommand{\eye}{{\mathrm{I}}}
\newcommand{\vecM}{{\mathrm{vec}}}

\newcommand{\Afnc}{\mathcal{A}}
\newcommand{\Bfnc}{\mathcal{B}}
\newcommand{\Cfnc}{\mathcal{C}}
\newcommand{\Dfnc}{\mathcal{D}}

\newcommand{\Kfnc}{\mathcal{K}}

\newcommand{\Dat}{\EuScript{D}_N}

\newcommand{\Dval}{\EuScript{D}_\mathrm{val}}

\newcommand{\futDat}[3]{\mathrm{#1}_{#3}^{#3+#2}}
\newcommand{\futDatc}[3]{\check{\mathrm{#1}}_{#3}^{#3+#2}}
\newcommand{\futDatcCor}[3]{\check{\mathrm{#1}}_{#3}^{#3+#2,(\corrFut)}}
\newcommand{\futDatt}[3]{\tilde{\mathrm{#1}}_{#3}^{#3+#2}}
\newcommand{\futDattCor}[3]{\tilde{\mathrm{#1}}_{#3}^{#3+#2,(\corrFut)}}
\newcommand{\futDath}[3]{\hat{\mathrm{#1}}_{#3|#3-1}^{#3+#2|#3+#2-1}}

\newcommand{\pastDatc}[3]{\check{\mathrm{#1}}_{#2}^{#2-#3}}
\newcommand{\pastDatt}[3]{\tilde{\mathrm{#1}}_{#2}^{#2-#3}}

\newcommand{\predDat}[4]{%
\ifthenelse{\equal{#4}{}}{%
\bar{\mathrm{#1}}_{#2}^{{\scriptscriptstyle(\mathrm{#3})}}%
}{%
\bar{\mathrm{#1}}_{#2}^{#4,{\scriptscriptstyle(\mathrm{#3})}}%
}%
}
\newcommand{\futWind}{\mathdutchcal{f}}
\newcommand{\pastWind}{\mathdutchcal{p}}
\newcommand{\corrFut}{\mathdutchcal{c}}

\newcommand{\HankStr}{\mathdutchcal{L}}

\newcommand{\initCond}[1]{\mathcal{X}_{#1}}
\newcommand{\initCondt}[1]{\tilde{\mathcal{X}}_{#1}}

\newcommand{\ACoef}{\mathdutchcal{A}}
\newcommand{\BCoef}{\mathdutchcal{B}}

\newcommand{\sing}{s}  % Singular value
\newcommand{\eig}{\lambda} % Eigenvalue

\makeatletter
\newcommand*\bigcdot{\mathpalette\bigcdot@{.5}}
\newcommand*\bigcdot@[2]{\mathbin{\vcenter{\hbox{\scalebox{#2}{$\m@th#1\bullet$}}}}}
\makeatother

\makeatletter
\newcommand{\pushright}[1]{\ifmeasuring@#1\else\omit\hfill$\displaystyle#1$\fi\ignorespaces}
\makeatother

\protected\edef\ell{\noexpand\ensuremath{{\mathchar\the\ell}}} % This fixes the 12th subequation problem of `autart`, see: http://tex.stackexchange.com/questions/285098/error-with-ifacconf-cls-and-subequations-with-more-than-twelve-equations?rq=1

\newcounter{ass_ch}
\newlist{assumption}{enumerate}{1}
\setlist[assumption]{label=\textbf{A.\arabic*}, ref={A.\arabic*},leftmargin={\widthof{A.99}},before={\setcounter{assumptioni}{\value{ass_ch}}}, after={\setcounter{ass_ch}{\value{assumptioni}}}}

\begin{document}

%% --------------------- arXiv.org title page for IEEE
%
%
\ifx\arXivPrint\undefined\else

\makeatletter
\twocolumn[{
\vspace{2cm}
This paper has been submitted to

\vspace{1cm}
\centerline{\textbf{\huge{Automatica}}}

\vspace{5cm}
%978-1-1234-5678-9/12/\$30.00 2020 IEEE \\
%\noindent\hspace*{0.5cm} DIO 10.1234/RADAR.2020.00011

\vspace{1cm}
\textbf{Citation}\\
P.B. Cox and R. T\'oth, ``Linear Parameter-Varying Subspace Identification: A Unified Framework ,'' %in \textit{}, pages ...-..., vol. no.  2020.

\vspace{1cm}
\textbf{Science Direct}\\
https://www.sciencedirect.com/science/article

\definecolor{commentcolor}{gray}{0.9}
\newcommand{\commentbox}[1] {\colorbox{commentcolor}{\parbox{\linewidth}{#1}}}

\vspace{4cm}
\commentbox{
	\vspace*{0.2cm}
	\hspace*{0.2cm}More papers from P. B. Cox can be found at\\~\\
	\centerline{\large{https://orcid.org/0000-0002-8220-7050}}
	\vspace*{0.0cm}
}
\commentbox{
	\vspace*{0.2cm}
	\hspace*{0.2cm}More papers from R. T\'oth can be found at\\~\\
	\centerline{\large{https://rolandtoth.eu/}}
	\vspace*{0.0cm}
}

\vspace{3cm}
\textcopyright 2020 Elsevier. Personal use of this material is permitted. Permission from Elsevier must be obtained for all other uses, in any current or future media, including reprinting/republishing this material for advertising or promotional purposes, creating new collective works, for resale or redistribution to servers or lists, or reuse of any copyrighted component of this work in other works.
}]
\clearpage
\makeatother

\fi
%
%
%% END: --------------------- arXiv.org Title page for IEEE

\begin{frontmatter}

\title{Linear Parameter-Varying Subspace Identification:\\A Unified Framework \vspace{-5mm}\thanksref{footnoteinfo}\thanksref{footnoteinfo2}} % Title, preferably not more 

\thanks[footnoteinfo]{This paper was not presented at any IFAC 
meeting. Corresponding author P.~B.~Cox. Tel. +31-40-2478188.}
\thanks[footnoteinfo2]{This paper has received funding from the European Research Council (ERC) under the European Union's Horizon 2020 research and innovation programme (grant agreement No 714663).}

\author[CSTUE]{Pepijn B. Cox}\ead{p.b.cox@tue.nl},    % Add the 
\author[CSTUE]{Roland T\'{o}th}\ead{r.toth@tue.nl},               % e-mail address 
%\author[Lille]{Mih\'{a}ly Petreczky}\ead{mihaly.petreczky@ec-lille.fr}%,  % (ead) as shown
%\author[Linkoping]{Lennart Ljung}\ead{ljung@isy.liu.se}

\address[CSTUE]{Control Systems Group, Department of Electrical Engineering, Eindhoven University of Technology, P.O. Box 513, \\ 5600 MB Eindhoven, The Netherlands.}  % Please supply                                              
          
                             % chosen from the IFAC 
										 % Five to ten keywords, 
										 % keyword list or with the 
                                          % help of the Automatica 
                                          % keyword wizard

\begin{abstract}                          % Abstract of not more than 200 words.
In this paper, we establish a unified framework for subspace identification (SID) of linear parameter-varying (LPV) systems to estimate LPV state-space (SS) models in innovation form. This framework enables us to derive novel LPV SID schemes that are extensions of existing linear time-invariant (LTI) methods. More specifically, we derive the open-loop, closed-loop, and predictor-based data-equations, an input-output surrogate form of the SS representation, by systematically establishing an LPV subspace identification theory. We show the additional challenges of the LPV setting compared to the LTI case. Based on the data-equations, several methods are proposed to estimate LPV-SS models based on a maximum-likelihood or a realization based argument. Furthermore, the established theoretical framework for the LPV subspace identification problem allows us to lower the number of to-be-estimated parameters and to overcome dimensionality problems of the involved matrices, leading to a decrease in the computational complexity of LPV SIDs in general. To the authors' knowledge, this paper is the first in-depth examination of the LPV subspace identification problem. The effectiveness of the proposed subspace identification methods are demonstrated and compared with existing methods in a Monte Carlo study of identifying a benchmark MIMO LPV system.\vspace{-4mm}
\end{abstract} \vspace{-4mm}
\end{frontmatter}

% ############################################### %
% ------------------------------------------------------------------------------ %
% 		Introduction
% ------------------------------------------------------------------------------ %
% ############################################### %

\section{Introduction} \vspace{-3mm}

Realization based state-space identification techniques, so-called \emph{subspace identification} (SID) methods, have been successfully applied in practice to estimate time-varying and/or nonlinear dynamical systems using \emph{linear parameter-varying} (LPV) \emph{state-space} (SS) models. Successful application examples range from diesel engines~\cite{Schulz2016}, wind-turbines~\cite{Wingerden2009,Felici2007a}, gas pipelines~\cite{Santos2010}, traffic flow models~\cite{Luspay2009}, and bioreactors~\cite{Verdult2002c} to nonlinear benchmark systems like the Lorenz attractor~\cite{Larimore2015}. The existing techniques are based on \emph{predictor based subspace identification} (PBSID)~\cite{Wingerden2009a}, \emph{past-output multivariable output-error state-space} (PO-MOESP)~\cite{Felici2007a}, \emph{canonical variate analysis} (CVA)~\cite{Larimore2015}, or the successive approximation identification algorithm~\cite{Santos2010}. However, these methods lack a common unified theory to tackle the LPV SID problem.

The field of subspace identification applies realization theory to find SS model estimates based on surrogate \emph{input-output} (IO) models (with appropriate noise models) directly estimated form data. These specialized IO models are estimated by using convex optimization and it can be shown that they correspond to a \emph{maximum-likelihood} (ML) estimate under the considered assumptions. Then, an SS realization is obtained from the IO model by either a \emph{direct realization} step or by an intermediate \emph{projection} step. In the latter idea, a projection is found to estimate the unknown state-sequence via matrix decomposition methods, then the SS matrices are estimated in a least-squares fashion. Obtaining such state-sequence is heavily based on realization theory, as the estimated state-basis should be consistent with the behavior of the underlying system. In the LTI setting, the IO model estimation and realization of the SS model under the presence of process and measurement noise is well understood~\cite{Overschee1996,Katayama2005,Verhaegen2007,Lindquist2015}. In the LPV case, contrary to the LTI setting, the stochastic interpretation of the methods with the appropriate noise representation is not well understood neither is the connection between the various methods have ever been studied.

LPV subspace schemes also suffer heavily from the curse of dimensionality, e.g., see~\cite[Table 1]{Wingerden2009a}, resulting  in ill-conditioning of the estimation problem and high parameter variance. Consequently, two common assumptions are taken to reduce the dimensionality: 
\begin{enumerate*}[label=(\roman*)]
	\item the excitation, in terms of the variation of the scheduling variable $p$, is periodic or white~\cite{Felici2007a,Wingerden2009}, and/or
	\item the output-equation of the SS representation is assumed to be $p$-independent~\cite{Schulz2016,Wingerden2009a,Luspay2009}. 
\end{enumerate*} 
However, such assumptions restrict practical applicability of the methods. To tackle ill-conditioning and to reduce estimation variance, kernel based regularization techniques have been proposed~\cite{Verdult2005,Wingerden2009a}. However, computational complexity of the involved kernels grows polynomially or exponentially w.r.t. the design parameters, which significantly compromises the effectiveness of these schemes.
Alternatively, SS models can directly be estimated by minimization of the $\ell_2$-loss in terms of the prediction-error associated with the model. These so-called \emph{prediction-error methods} (PEM) minimize the $\ell_2$-loss directly using gradient-based methodologies~\cite{Lee1997a,Verdult2003,Wills2008,Wills2011} or by the expectation-maximization scheme~\cite{Wills2011}. However, minimization of the $\ell_2$-loss w.r.t. to the LPV-SS model parameters is a nonlinear and nonunique optimization problem, requiring an initial estimate close to the global optimum.

The goal of this paper is to obtain a unified formulation to treat the LPV subspace identification problem and derive its associated stochastic properties by systematically establishing an LPV SID theory. This unified framework enables us to
\begin{enumerate*}[label=(\roman*)]
	\item understand relations and performance of LPV SIDs,
	\item extend most of the successful LTI subspace schemes to the LPV setting, 
	\item decrease the dimensionality problems, and 
	\item relax assumptions on the scheduling signal. 
\end{enumerate*}
In addition, we establish stochastic LPV realization theory which provides state estimation with maximum likelihood efficiency. To the authors' knowledge, this paper is the first in-depth treatment of the subspace theory in the LPV case. In this paper, we focus on projection based schemes, but the direct realization schemes can easily be abstracted from the developed results, i.e., see~\cite{Cox2018PHD}. We follow well-known concepts from the LTI literature, e.g.,~\cite{Overschee1996,Verhaegen2007}, and our theoretic results are also based on preliminary studies in the LPV setting~\cite{Verdult2002a,Verdult2005,Wingerden2009a}. The main contributions of this paper are: \vspace{-2mm}
\begin{enumerate}[label=\roman*)]
	\item Formulating the state estimation problem by a maximum-likelihood approach based on canonical correlation analysis and by a realization based approach.
	\item Stochastic interpretation of state estimation with maximum likelihood efficiency under the presence of noise.
	\item Computationally efficient formulation of SIDs to decrease the effects of the curse of dimensionality.
\end{enumerate}\vskip -5mm \noindent

The unified subspace theory is tackled in the global identification setting, i.e., under general trajectories of the scheduling signal, contrary to some results in the literature \cite{Felici2007a,Wingerden2007,Santos2008}. %\footnote{\cite{Santos2008} initializes the recursions for its global identification methodology by subspace scheme in the local setting.} . 

The proposed schemes could also be applied in a setting where the scheduling signal contains additive white noise that might be correlated to the input additive noise. In such case, the IO estimation step could be performed by using an instrumental variable approach~\cite{Toth2012b}. However, investigation of such formulation is outside of the scope of this paper.

This paper is organized as follows: first, the assumed data-generating system with LPV-SS representation and general innovation noise structure are presented and the open-loop, closed-loop, and predictor-based data-equations are derived (Sec.~\ref{sec:data-equations}). Then, the considered parametric LPV-SS identification problem is introduced (Sec.~\ref{sec:ParamSID}). Next, the state realization problem is tackled from a maximum-likelihood and realization based argument first for the open-loop identification setting (Sec.~\ref{sec:SIDOLTimeInv}) and then for the closed-loop identification setting (Sec.~\ref{sec:SIDCLTimeInv}) leading to the LPV formulation of various well-known LTI subspace methods. %To lower a computational demand, the basis reduced formulation is introduced (Sec.~\ref{sec:BasisReduced}). 
The efficiency of the unified framework is demonstrated by a Monte Carlo study on an LPV-SS identification benchmark (Sec.~\ref{sec:SimulationExample}).

% ############################################### %
% ------------------------------------------------------------------------------ %
% 		The LPV data-equations
% ------------------------------------------------------------------------------ %
% ############################################### %
\vspace{-3mm}
\section{The LPV data-equations} \label{sec:data-equations}\vspace{-3mm}
In this section, surrogate input-output representations of SS models are formulated which are key in solving the subspace identification problem. Namely, we derive the LPV open-loop data-equation (Sec.~\ref{subsec:OLdataEq}), closed-loop data-equation (Sec.~\ref{subsec:CLdataEq}), and the predictor-based data-equation (Sec.~\ref{subsec:predictorEq}) for LPV data-generating systems in a SS form (Sec.~\ref{subsec:data-generating}).

% ####################################################
% Subsec: The data-generating system
% ####################################################
\vspace{-2mm}
\subsection{The data-generating system} \label{subsec:data-generating}\vspace{-2mm}

The goal is to obtain an SS model estimate of the data-generating system $\sys_\mathrm{o}$ represented in the following LPV-SS \emph{innovation form}\footnote{In the majority of the subspace literature~\cite{Overschee1996,Verhaegen2007,Wingerden2009}, the data-generating system is assumed to be in the innovation form as given in~\eqref{eq:SSrep}. However, in~\cite{Cox2018PHD}, it is shown that the noise description in~\eqref{eq:SSrep} is not equivalent to a state-space form with general noise representation, i.e., a representation with different noise processes on the state and output equation. \cite{Cox2018PHD} also shows that a static, affine $\Kfnc(p_t)$ can approximate the general setting if the state dimension is increased. In practice, we often need to restrict parameterization of $\Kfnc$, e.g., to the static, affine parameterization in~\eqref{eq:SSrep}, to reduce complexity of the estimation method and variance of the model estimates. Hence, despite the possible increase of the state order of the equivalent innovation form, the usage of this affine form has been found adequate in practical applications~\cite{Luspay2009,Wingerden2009,Felici2007a,Schulz2016}.}\vspace{-2mm}
\begin{subequations} \label{eq:SSrep}
\begin{alignat}{3}
  x_{t+1} &= \Afnc(p_t)&x_t&+\Bfnc(p_t)&&u_t+\Kfnc(p_t)\xi_t, \label{eq:SSrepState} \\
  y_t &= \Cfnc(p_t)&x_t&+\Dfnc(p_t)&&u_t+\xi_t, \label{eq:SSrepOut}
\end{alignat}
\end{subequations} \vskip -5mm \noindent
where $x:\sZ\rightarrow\sX=\Rx$ is the state variable, $y:\sZ\rightarrow\sY=\Ry$ is the measured output signal, $u:\sZ\rightarrow\sU=\Ru$ is the input signal, $p:\sZ\rightarrow\sP\subset\Rp$ is the scheduling signal, and $\xi:\sZ\rightarrow\Ry$ is the sample path realization of the zero-mean stationary process: \vspace{-4mm}
\begin{equation} \label{eq:noisy}
\boldsymbol \xi_t \sim \mathcal{N}(0,\Xi^2),
\end{equation} \vskip -5mm \noindent
where $\boldsymbol \xi_t:\Omega\rightarrow\Ry$ is a white noise process with sample space $\Omega$ (set of possible outcomes) and $\Xi^2\in\mathbb{R}^{\NX\times\NX}$ is a positive definite covariance matrix. Furthermore, we will assume $u,p,y,\xi$ to have left compact support to avoid technicalities with initial conditions. The matrix functions $\Afnc(\cdot),\ldots,\Kfnc(\cdot)$ defining the SS representation \eqref{eq:SSrep} are affine combinations of bounded scalar functions $\psi^\ind{i}(\cdot):\sP\rightarrow\mathbb{R}$: \vspace{-3mm}
\begin{equation}\label{eq:sysMatrices}
\begin{aligned} 
 \Afnc(p_t)&\!=\! \sum_{i=0}^{\NPSI} A_i\psi^\ind{i}(p_t),\hspace{-1mm}&\Bfnc(p_t)&\!=\!\hspace{-1mm}\sum_{i=0}^{\NPSI} B_i\psi^\ind{i}(p_t) , \\
 \Cfnc(p_t)&\!=\!\sum_{i=0}^{\NPSI} C_i\psi^\ind{i}(p_t),\hspace{-1mm}&\Dfnc(p_t)&\!=\!\hspace{-1mm}\sum_{i=0}^{\NPSI} D_i\psi^\ind{i}(p_t), \\
 \Kfnc(p_t)&\!=\!\sum_{i=0}^{\NPSI} K_i\psi^\ind{i}(p_t), 
\end{aligned}
\end{equation}\vskip -7mm \noindent
where $\{A_i,B_i,C_i,D_i,K_i\}_{i=0}^{\NPSI}$  are constant, real matrices with appropriate dimensions and $\psi^\ind{0}(\cdot)=1$ is assumed to be constant. Additionally, for well-posedness, it is assumed that $\{\psi^\ind{i}\}_{i=1}^{\NPSI}$ are  linearly independent over an appropriate function space and are normalized w.r.t. an appropriate norm or inner product~\cite{Toth2012}. Due to the freedom to consider arbitrary functions $\psi^\ind{i}$, \eqref{eq:sysMatrices} can capture a wide class of static nonlinearities and time-varying behaviors. For notational simplicity, we define $\psi_t = [ \begin{array}{ccc} \psi^\ind{0}(p_t) & \ldots&\psi^\ind{\NPSI}(p_t) \end{array} ]$.

% ####################################################
% Subsec: The open-loop data-equation
% ####################################################
\vspace{-2mm}
\subsection{The open-loop data-equation} \label{subsec:OLdataEq} \vspace{-2mm}

The first step in tackling the subspace identification problem is to represent the dynamics of the data-generating system~\eqref{eq:SSrep} as an equivalent IO representation, the so-called \emph{data-equation}. The unknowns in these data-equations are estimated by convex optimization and the final SS model is obtained from these data-equations using matrix decomposition techniques (see Sec.~\ref{sec:ParamSID}-\ref{sec:SIDCLTimeInv} for more details). Hence, the data-equations are key in formulating the subspace problem. 

Open-loop data-equations are rarely used in the literature, as the innovation noise $\xi_t$ is unknown. In light of the MAX identification setting in \cite{Cox2016c,Cox2018PHD}, the innovation noise $\xi_t$ can be uniquely obtained by convex optimization, which renders the open-loop equations attractive for further investigation, similar to~\cite{Mercere2016} in the LTI setting.
Using~\eqref{eq:SSrepOut}, the output w.r.t.\ a \emph{future window} $\futWind\in\mathbb{N}_{+}$, where $\mathbb{N}_{+}=\{i\in\mathbb{Z}~\vert~ i>0\}$, starting from time-instance $t$ can be written as \vspace{-2mm}
\begin{equation} \label{eq:FutOutOL}
\futDat{y}{\futWind}{t} = (\obsvTV_\futWind\diamond p)_t x_t + (\check\HankStr_\futWind \diamond p )_t \futDatc{z}{\futWind}{t} + \futDat{\xi}{\futWind}{t},
\end{equation} \vskip -5mm \noindent
where $\check z_t=[u^\top_t~\xi^\top_t]$ is the extended ``input'' signal and $\futDat{y}{\futWind}{t}$, $\futDat{\xi}{\futWind}{t}$, and $\futDatc{z}{\futWind}{t}$ are sequences according to the notation \vspace{-2mm}
\begin{equation*}
\mathrm{q}_l^s = \left\{ \begin{array}{ll} 
	\bigl[ \begin{array}{cccc} q^\top_l & q^\top_{l+1} & \cdots  & q^\top_{s-1}  \end{array} \bigr]^\top & \mbox{ if } s > l, \\  
	\bigl[ \begin{array}{cccc} q^\top_{l-1} & \cdots  & q^\top_{s+1} & q^\top_s  \end{array} \bigr]^\top & \mbox{ if } s < l. \end{array}   \right.
\end{equation*} \vskip -6mm \noindent
Furthermore, the matrix functions in~\eqref{eq:FutOutOL} are given as \vspace{-3mm}
\begin{subequations}
\begin{align}
(\obsvTV_\futWind \diamond p)_t &= \left[\! \begin{array}{ccc} \Cfnc^\top\!(p_t) & \!\cdots\!\! & \Bigl(\Cfnc(p_{t+\futWind})\!\! \prod\limits_{i=1}^\futWind \!\!\Afnc(p_{t+f-i})\Bigr)^{\!\top} \end{array}\! \right]^{\!\top}\!\!\!, \label{eq:obsvInit} \\
\check \Bfnc (p_t) &= \left[ \begin{array}{cc} \Bfnc(p_t) & \Kfnc (p_t)\end{array} \right], \\
\check \Dfnc(p_t) &= \left[ \begin{array}{cc} \Dfnc(p_t) & 0_{\NY\times\NY}\end{array} \right], 
\end{align} \vskip -6mm \noindent
and $\check\HankStr_\futWind$ is as given in~\eqref{eq:HankShift} where $\Afnc_t,\ldots,\check\Dfnc_t$ is a shorthand notation for  $\Afnc(p_t),\ldots,\check\Dfnc(p_t)$. Here, $\prod_{i=1}^\futWind$ is considered with left multiplication.
\begin{figure*}[!ht]
%The matrix function $\check\HankStr_\futWind$ is given by \vspace{-2mm}
\begin{equation} \label{eq:HankShift}
(\check\HankStr_\futWind\diamond p)_t =  \left[ \begin{array}{ccccc}
\check\Dfnc_t & 0 & 0 & \cdots & 0 \\
\Cfnc_{t+1}\check\Bfnc_t & \check\Dfnc_{t+1} & 0 & \cdots & 0 \\
\Cfnc_{t+2}\Afnc_{t+1}\check\Bfnc_t &\Cfnc_{t+2}\check\Bfnc_{t+1} & \check\Dfnc_{t+2} & \cdots & 0 \\
\vdots & \vdots & \ddots & \ddots & \vdots \\
\Cfnc_{t+\futWind-1}\!\!\left[\prod\limits_{i=2}^{\futWind-1}\! \!\Afnc_{t+\futWind-i}\right]\!\!\check\Bfnc_t &  ~\Cfnc_{t+\futWind-1}\!\!\left[\prod\limits_{i=2}^{\futWind-2} \!\! \Afnc_{t+\futWind-i}\right]\!\!\check\Bfnc_{t+1} & ~\Cfnc_{t+\futWind-1}\!\!\left[\prod\limits_{i=2}^{\futWind-3}\!\! \Afnc_{t+\futWind-i}\right]\!\!\check\Bfnc_{t+2} & \cdots &  \check\Dfnc_{t+\futWind-1}\!\! \end{array} \right]\!
\end{equation} %\vskip -3mm %\noindent
%where $\Afnc_t,\ldots,\check\Dfnc_t$ is a shorthand notation for  $\Afnc(p_t),\check\Bfnc(p_t),\Cfnc(p_t),\check\Dfnc(p_t)$. 

\hrulefill
\end{figure*}
\end{subequations}
In~\eqref{eq:FutOutOL} and~\eqref{eq:obsvInit}-\eqref{eq:HankShift},  the $\diamond$ operator is a shorthand notation for dynamic dependency on the scheduling signal, i.e., $(\obsvTV_\futWind \diamond p)_t = \obsvTV_\futWind (p_t,p_{t-1},p_{t-2},\ldots)$.

Next, the state can be decomposed by using the past values of the input and noise signals: \vspace{-3mm}
\begin{equation}  \label{eq:StateOL}
x_t = ( \check\reachTV_\pastWind \diamond p)_t \pastDatc{z}{t}{\pastWind} + \initCond{\pastWind},
\end{equation} \vskip -6mm \noindent
with \emph{past window} $\pastWind\in\mathbb{N}_{+}$, past data $\pastDatc{z}{t}{\pastWind}$, and \vspace{-3mm}
\begin{subequations}
\begin{align}
(\check\reachTV_\pastWind \diamond p)_t &= \biggl[\!\begin{array}{cccc} \check\Bfnc(p_{t-1})& \Afnc(p_{t-1})\check\Bfnc(p_{t-2})&\cdots \end{array} \nonumber \\
 &\hspace{2.5cm}\left[ \prod\limits_{i=1}^{\pastWind-1}\!\!\Afnc(p_{t-i}) \right]\check\Bfnc(p_{t-\pastWind})\biggr], \label{eq:TVReach} \\
 \initCond{\pastWind} &= \left[ \prod_{i=1}^\pastWind \Afnc(p_{t-i}) \right]x_{t-\pastWind}. \label{eq:InitCondOL}
\end{align}\vskip -6mm \noindent
%
%
%with the initial condition
%
%\begin{equation}
%\initCond{\pastWind} = \left[ \prod_{i=1}^\pastWind \Afnc(p_{t-i}) \right]x_{t-\pastWind}.
%\end{equation}
%
%
%The past data is given as
%
%
%\begin{equation} \label{eq:pastData}
%\futDatc{z}{t}{\pastWind} = \left[  \begin{array}{ccccc} u^\top_{t-1} & \xi^\top_{t-1} & \hdots & u^\top_{t-\pastWind} & \xi^\top_{t-\pastWind} \end{array} \right]^\top.
%\end{equation}
\end{subequations}
%
%
%Remark the subtle difference in notation between the \emph{future} data vector $\futDatc{z}{\futWind}{t}$ and \emph{past} data vector $\pastDatc{z}{t}{\pastWind}$ in~\eqref{eq:FutOutOL} and~\eqref{eq:StateOL}.
%
Combining the output-equation based on the future values~\eqref{eq:FutOutOL} with the state-equation based on the past values~\eqref{eq:StateOL} results in the \emph{open-loop data-equation} \vspace{-2mm}
\begin{multline} \label{eq:DataEqOL}
\futDat{y}{\futWind}{t} = (\obsvTV_\futWind  \check\reachTV_\pastWind \diamond p)_t \pastDatc{z}{t}{\pastWind} + (\check\HankStr_\futWind \diamond p)_t  \futDatc{z}{\futWind}{t} + \futDat{\xi}{\futWind}{t} \\ + (\obsvTV_\futWind\diamond p)_t\initCond{\pastWind},
\end{multline} \vskip -5mm \noindent
which has the form of a MIMO LPV-IO model. Estimating the underlying IO relationship of~\eqref{eq:DataEqOL} requires the input-scheduling pair $(u,p)$ and the innovation noise $\xi$ to be uncorrelated in order to obtain an unbiased estimate of the relationship~\eqref{eq:DataEqOL} under PEM, e.g., see \cite{Jansson2005,Chiuso2007a,Verhaegen2007}. The case when $(u,p)$ and $\xi$ are uncorrelated is usually referred to as the open-loop identification setting~\cite{Eykhoff1974,Ljung1999,Verhaegen2007}, characterized by the following two assumptions: \vspace{-2mm}
\begin{assumption}
	\item\label{ass:ueuncorrOL} The input signal $u$ is quasi-stationary  and uncorrelated with $\xi$, i.e., $\bar\expct\{u_t (\xi_{t+\tau})^\top\}=\bar\expct\{u_t (\xi_{t-\tau})^\top\}=0$ for all $\tau\in\mathrm{N}_0$. %
\footnote{The generalized expectation operation $\bar\expct$ of a process $u$ is defined as $\bar\expct\left\{u_t \right\} = \lim_{N\rightarrow\infty} \frac{1}{N} \sum_{t=1}^{N} \expct\{u_t\}$. A process $ u$ is said to be \emph{quasi-stationary} if there exists finite $c_1,c_2\in\mathbb{R}$ such that
\begin{enumerate*}[label=\roman*)]
	\item $\left\|\expct\{u_t \}\right\|_2 < c_1$ for all $t$, and
	\item $\left\|\Tr\bigl(\bar\expct\{ u_t u^\top_{t-\tau} \}\bigr)\right\|_2 < c_2$ for all $\tau$, e.g., see~\cite{Ljung1999}. 
\end{enumerate*}}
	\item\label{ass:peuncorrOL} The scheduling signal $p$ is quasi-stationary and uncorrelated with $\xi$.
\end{assumption} \vskip -2mm \noindent
Assumptions~\ref{ass:ueuncorrOL} and~\ref{ass:peuncorrOL} are not restricting, for example, when considering the LPV modeling problem of a thermal loop in a wafer scanner. The thermal distribution of the wafer varies with the position, but it does not influence the measurement noise of the position sensor and, therefore, the position as scheduling signal fulfills Assumption~\ref{ass:peuncorrOL}. On the other hand, an inverted pendulum setup with stabilizing controller where the angle of the pendulum is the scheduling signal (and output) will not satisfy Assumptions~\ref{ass:ueuncorrOL}-\ref{ass:peuncorrOL}. In such a case, $p$ is correlated with past values of $\xi$ due to the closed-loop interconnection between plant and controller.

%Assumption~\ref{ass:ueuncorrOL} coincides with the majority of literature on LTI system identification, e.g.,~\cite{Eykhoff1974,Ljung1999,Verhaegen2007,Pintelon2012}, while Assumption~\ref{ass:peuncorrOL} is an additional requirement of the LPV case.

% ####################################################
% Subsec: The closed-loop data-equation
% ####################################################
\vspace{-2mm}
\subsection{The closed-loop data-equation} \label{subsec:CLdataEq} \vspace{-2mm}

To overcome the limitations of the open-loop setting, the data-equation~\eqref{eq:DataEqOL} can be written in an alternative form. Analogously to LTI identification ~\cite{Jansson2005,Chiuso2007a,Verhaegen2007}, the output-equation~\eqref{eq:SSrepOut} is substituted in the state-equation~\eqref{eq:SSrepState}, resulting in \vspace{-3mm}
\begin{equation} 
	x_{t+1} = \tilde\Afnc(p_t) x_t+ \tilde\Bfnc(p_t)\tilde z_t,  \label{eq:SSrepInnStateCL}
\end{equation} \vskip -5mm \noindent
where $\tilde z_t = [ u_t^\top~ y_t^\top ]^\top$ and the corresponding matrix functions are \vspace{-2mm}
\begin{subequations} \label{eq:SSMatStateCL}
\begin{align}
\tilde\Afnc(p_t) &= \Afnc(p_t) \!-\! \Kfnc(p_t) \Cfnc(p_t), \\
\tilde\Bfnc(p_t) &= [\begin{array}{cc} \Bfnc(p_t) \!-\! \Kfnc(p_t) \Dfnc(p_t)& ~\Kfnc(p_t) \end{array}].
\end{align} \vskip -6mm \noindent
\end{subequations}
It is important to note that~\eqref{eq:SSrepInnStateCL} does not depend explicitly on the stochastic process $\xi$. Hence, the state-equation~\eqref{eq:SSrepInnStateCL} can be treated in a deterministic setting. However, moving from the open-loop to the closed-loop dynamics comes at the cost of polynomial dependency of the $\tilde\Afnc$ and $\tilde\Bfnc$ matrix functions. Opposed to the LTI setting where $\Kfnc=K\in\mathbb{R}^{\NX\times\NY}$, applying~\eqref{eq:SSrepInnStateCL} instead of~\eqref{eq:SSrepState} increases the model complexity. Using~\eqref{eq:SSrepInnStateCL}, the stacked output-equation~\eqref{eq:FutOutOL} can equivalently be represented as \vspace{-3mm}
\begin{equation} \label{eq:FutOutCL}
\futDat{y}{\futWind}{t} = (\tilde\obsvTV_\futWind\diamond p)_t x_t + (\tilde\HankStr_\futWind\diamond p)_t \futDatt{z}{\futWind}{t} +  \futDat{\xi}{\futWind}{t}.
\end{equation}\vskip -6mm \noindent
In~\eqref{eq:FutOutCL}, $(\tilde\obsvTV_\futWind\diamond p)_t$ denotes the observability matrix with $\tilde\Afnc$ instead of $\Afnc$, $(\tilde\HankStr_\futWind\diamond p)_t$ is constructed with $\tilde\Afnc$, $\tilde\Bfnc$, and the future values $\futDatt{z}{\futWind}{t}$ are similarly stacked as $\futDatc{z}{\futWind}{t}$ in~\eqref{eq:FutOutOL}. Note that $\futDatc{z}{\futWind}{t}$ is dependent on the pair $(u_t,\xi_t)$ and $\futDatt{z}{\futWind}{t}$ on $(u_t,y_t)$. The state can be written as a combination of past signals (similar to~\eqref{eq:StateOL}) \vspace{-4mm}
\begin{equation} \label{eq:StateCL}
x_t = ( \tilde\reachTV_\pastWind \diamond p)_t \pastDatt{z}{t}{\pastWind} + \initCondt{\pastWind},
\end{equation}\vskip -6mm \noindent
where $\pastWind\in\mathbb{N}_{+}$ is the past window, $(\tilde\reachTV_\futWind\diamond p)_t$ denotes the reachability matrix~\eqref{eq:TVReach} with $\tilde\Afnc$ and $\tilde\Bfnc$ instead of $\Afnc$ and $\Bfnc$, and $\initCondt{\pastWind}$ is the initial condition~\eqref{eq:InitCondOL} with $\tilde\Afnc$ instead of $\Afnc$.

Combining~\eqref{eq:FutOutCL} and~\eqref{eq:StateCL} results in the \emph{closed-loop data-equation}: \vspace{-3mm}
\begin{multline}  \label{eq:DataEqCL}
\futDat{y}{\futWind}{t} = (\tilde\obsvTV_\futWind \tilde\reachTV_\pastWind \diamond p)_t \pastDatt{z}{t}{\pastWind} + (\tilde\HankStr_\futWind\diamond p)_t \futDatt{z}{\futWind}{t} +  \futDat{\xi}{\futWind}{t} \\ + (\tilde\obsvTV_\futWind\diamond p)_t \initCondt{\pastWind}.
\end{multline}\vskip -8mm \noindent

To formulate our identification problem in the closed-loop case, we take the following assumptions: \vspace{-2mm}
\begin{assumption}
	\item\label{ass:ueuncorrCL} The input signal $u$ is quasi-stationary and uncorrelated with future values of $\xi$, i.e., $\bar\expct\{u_t (\xi_{t+\tau})^\top\}=0$ for all $\tau\in\mathbb{N}_0$.
	\item\label{ass:peuncorrCL} The scheduling signal $p$ is quasi-stationary and uncorrelated with future values of $\xi$.
\end{assumption}\vskip -2mm \noindent
Assumptions~\ref{ass:ueuncorrCL} and~\ref{ass:peuncorrCL} allow to identify systems under general feedback structures, e.g., see~\cite{Eykhoff1974,Ljung1999,Verhaegen2007}.

% ####################################################
% Subsec: Derivation of the predictor
% ####################################################
\vspace{-2mm}
\subsection{Derivation of the predictor} \label{subsec:predictorEq} \vspace{-2mm}

A commonly applied data-equation for subspace identification is the predictor form, e.g., see~\cite{Chiuso2005,Chiuso2007a,Wingerden2009a}. %The subspace method employ a two-sided expansion form that can be seen as an infinite order ARX representation, which gives a predictor of the underlying IO model relationship. To obtain the one-step-ahead predictor of the state, we will remove the innovation noise from the state-equation. To this end, by applying the closed-loop state dynamics, the one-step-ahead predictor of the state based on the past window $\pastWind$ is
%
%\begin{multline*}
%\hat x_{t|t-1} =  \left[ \prod_{i=1}^\pastWind (\tilde\Afnc \diamond p)_{t-i}\right]\check x_{t-\pastWind} \\ 
%+ \sum_{i=1}^\pastWind\left[ \prod_{j=1}^{i-1} (\tilde\Afnc \diamond p)_{t-j}\right] (\tilde\Bfnc\diamond p)_{t-i} \tilde z_{t-i}.
%\end{multline*}
%
To formulate the one-step-ahead predictor for the output, the closed-loop state~\eqref{eq:SSrepInnStateCL} is substituted into the output-equation~\eqref{eq:SSrepOut} and we take the conditional expectation, resulting in: \vspace{-4mm}
\begin{multline} \label{eq:yPredIO}
\hat y_{t|t-1} = \Cfnc(p_t)  \initCondt{\pastWind} + \Dfnc(p_t) u_t \\ 
+  \sum_{i=1}^\pastWind\Cfnc(p_t)  \left[ \prod_{j=1}^{i-1} (\tilde\Afnc \diamond p)_{t-j}\right] (\tilde\Bfnc\diamond p)_{t-i} \tilde z_{t-i} .
\end{multline}\vskip -7mm \noindent
Note that~\eqref{eq:yPredIO} is the minimal variance estimator of $y_t$ and that~\eqref{eq:yPredIO} represents an LPV-ARX model where $\pastWind\rightarrow\infty$ will diminish the influence of the initial condition $\initCondt{\pastWind}$ under the assumption that $\tilde\Afnc$ is stable. The one-step-ahead predictor of the output can be similarly stacked as the closed-loop data-equation~\eqref{eq:DataEqCL} leading to the \emph{predictor-based data-equation}: \vspace{-4mm}
\begin{multline}  \label{eq:DataEqPred}
\futDath{y}{\futWind}{t} = (\tilde\obsvTV_\futWind \tilde\reachTV_\pastWind \diamond p)_t\pastDatt{z}{t}{\pastWind} + (\tilde\HankStr_\futWind \diamond p)_t \futDatt{z}{\futWind}{t} + \\
(\tilde\obsvTV_\futWind\diamond p)_t  \initCondt{\pastWind}.
\end{multline}\vskip -6mm \noindent
Note that~\eqref{eq:DataEqPred} is the one-step-ahead predictor of~\eqref{eq:DataEqCL}. Hence, the SS representation of $\sys_\mathrm{o}$ can be captured by the predictor~\eqref{eq:yPredIO} from which~\eqref{eq:DataEqPred} can be constructed~\cite{Chiuso2005,Chiuso2007a,Wingerden2009,Wingerden2009a}.

\vspace{-2mm}
\section{Parametric subspace identification setting} \label{sec:ParamSID} \vspace{-2mm}

Known LTI and LPV subspace schemes are based on the aforementioned data-equations or their simplifications. The subspace schemes rely on matrix decomposition techniques on applied $\obsvTV_\futWind\reachTV_\pastWind$ to obtain a realization of these two matrices; however, these decomposition techniques cannot be directly applied to parameter-varying matrices. As shown in~\cite{Wingerden2009a}, the main difficulty comes from the time-varying observability matrix, as the dependency structure of the reachability matrix can be absorbed in an extended input vector.

In this paper, we are interested in estimating the unknown matrices $\{A,\ldots,K_i\}_{i=0}^\NPSI$ corresponding to parameters $\theta_{A}=[\  \vecM\{A_0\}^\top \  \cdots \ \  \vecM\{A_\NPSI\}^\top\ ]^\top$. The collection of unknown parameters is denoted by  $\theta=[\ \theta^\top_{A} \ \ \cdots \ \ \theta^\top_{K}\ ]^\top$ with $\theta\in\Theta=\mathbb{R}^\NTH$ and $\NTH=(1+\NPSI)(\NXp[2]+2\NY\NX+\NU\NX+\NY\NU)$. The parameters of the data-generating system $\sys_\mathrm{o}$ are denoted as $\theta_\mathrm{o}$ and we denote with $\sys(\theta')$ the model~\eqref{eq:SSrep} with parameters $\theta'$. The identification problem of SS models based on a \emph{data set} $\Dat=\{(y_t,p_t,u_t)\}_{t=1}^N$ has non-unique solutions up to a transformation matrix, e.g., see~\cite{Verdult2002a,Cox2018PHD}. Hence, we aim at identifying an isomorphic, jointly state minimal $\sys(\theta)$ w.r.t. $\sys(\theta_0)$ defined by the following set: %
\footnote{The representation $\sys$ is jointly state minimal if $\check\reachTV_\NX$ and $\obsvTV_\NX$ have at least $\NX$ linearly independent rows or columns, respectively, in a function sense, i.e., $\rank(\check\reachTV_\NX)=\NX$ and $\rank(\obsvTV_\NX)=\NX$.}
\vspace{-4mm}\begin{multline} \label{eq:setOfParametersdetReal}
\mathcal{I}_{\theta}=\Big\{~ \theta^\prime ~~\Big|~~ \exists T\in\mathbb{R}^{\NX\times\NX} \mbox{ s.t. } \rank(T)=\NX \\[-1mm] \mbox{ and } \theta^\prime = S(\theta,T) \Big\},
\end{multline}\vskip -7mm \noindent
where the indistinguishable manifold $S$ is given in~\eqref{eq:indistMan}.

\begin{figure*}[!ht]
%The indistinguishable manifold $S$ in~\eqref{eq:setOfParametersdetReal} is defined as 
\vspace{-4mm}
\begin{multline} \label{eq:indistMan}
S(\theta,T) \!=\! \big[\! \begin{array}{ccccccccc}
\vecM\{T^{-1}A_0T\}^{\!\top} &  \!\!\cdots\!\! & \vecM\{T^{-1}A_\NPSI T\}^{\!\top} & \vecM\{T^{-1}B_0\}^{\!\top} & \!\!\cdots\!\! & \vecM\{T^{-1}B_\NPSI\}^{\!\top} & \vecM\{C_0T\}^{\!\top} & \!\!\cdots\!\!  & \vecM\{C_\NPSI T \}^{\!\top}  \end{array} \\
\begin{array}{cccccc} \vecM\{D_0\}^{\!\top} &  \!\!\cdots\!\! & \vecM\{D_\NPSI\}^{\!\top} & \vecM\{T^{-1}K_0\}^{\!\top} &  \!\!\cdots\!\! & \vecM\{T^{-1}K_\NPSI\}^{\!\top} \!\!\end{array} \big]^{\!\top}\!.
\end{multline}\vskip -5mm \noindent
\hrulefill
\end{figure*}

Given a data set $\Dat$ and the basis functions $\{\psi^\ind{i}\}_{i=0}^\NPSI$, the goal of this paper is obtain a consistent estimate $\hat\theta$ of the data-generating system $\sys_\mathrm{o}$ such that $\hat\theta\rightarrow\theta\in\mathcal{I}_{\theta_\mathrm{o}}$ with probability one as $N\rightarrow\infty$. For the identification setting to be well-posed, the following standard assumptions are taken \vspace{-2mm}
\begin{assumption}
	\item\label{ass:datInM} $\sys(\theta_\mathrm{o})$ is an element of the model set, meaning that $\exists \theta\in\Theta$ such that $\theta\in\mathcal{I}_{\theta_\mathrm{o}}$.
	\item\label{ass:fullObsvCont} The state-minimal SS representation with static, basis affine dependency structure of the system $\sys_\mathrm{o}$ is structural state-observable w.r.t.\ to the pair $(\Afnc(p_t),\Cfnc(p_t))$ and structurally state-reachable w.r.t.\ to the pair$(\Afnc(p_t),\allowbreak[\Bfnc(p_t)~\Kfnc(p_t)\Xi^{-1}])$ \cite[Lem. 2.4]{Cox2018PHD}.
	\item\label{ass:stabilitySys} The open-loop dynamics $\Afnc(p_t)$ or closed-loop dynamics $\tilde \Afnc(p_t)$ are asymptotically stable for the open-loop or closed-loop cases, respectively.
	\item\label{ass:initialCond} The past window $\pastWind$ is chosen sufficiently large, such that $\initCond{\pastWind}\approx 0$ or $\initCondt{\pastWind}\approx 0$, $\forall p\in\sP^\sZ$ for the open-loop or closed-loop cases, respectively.
\end{assumption}\vskip -2mm \noindent
We can only estimate system dynamics that manifest in the data, so the system is represented with a structurally minimal IO equivalent SS representations, as formalized in Assumption~\ref{ass:fullObsvCont}. With Assumption~\ref{ass:stabilitySys}, the influence of the initial state $x_{t-\pastWind}$ can be neglected in~\eqref{eq:DataEqOL}, \eqref{eq:DataEqCL}, or \eqref{eq:DataEqPred}. This property is widely applied in subspace identification~\cite{Overschee1996,Verdult2002b,Jansson2003,Chiuso2005,Wingerden2009a}. See~\cite[Lemma 5]{Verdult2002b} for an upper-bound on the approximation error of this assumption.
Note that, we do not take the assumption that either $(\Cfnc(p),\Dfnc(p))$ or $\Kfnc(p)$ are parameter independent to reduce the complexity of the IO model opposed to state-of-the-art subspace schemes~\cite{Verdult2005,Wingerden2009a}.

Next, we will develop a unified theory to extend the LTI N4SID, MOESP, CVA, SS-ARX, and PBSID principles to the LPV case. There are two significant differences with respect to the LTI case. Firstly, almost all LTI formulations apply a (partial) ARX model structure, however, in the LPV case, the LPV-ARX model comes with significantly larger parameterization compared to the MAX representation in the open-loop setting. Secondly, we apply a predictor pre-estimation step to identify the unknown quantities of the matrices $\obsvTV_\futWind\check\reachTV_\pastWind$, $\check\HankStr_\futWind$, $\tilde\obsvTV_\futWind\tilde\reachTV_\pastWind$, etc. and construct the full matrices instead of estimating the matrices $\obsvTV_\futWind\check\reachTV_\pastWind$, $\check\HankStr_\futWind$, $\tilde\obsvTV_\futWind\tilde\reachTV_\pastWind$ directly using the data-equations~\eqref{eq:DataEqOL} or~\eqref{eq:DataEqCL}. Direct estimation of the matrices by oblique projections comes with a significant computational cost \cite[Table 1]{Verdult2002b} compared to the predictor formulation \cite[Table 1]{Wingerden2009a}, especially in the LPV case. 
Furthermore, direct estimation of the matrices will not take the structural restrictions of $\check\HankStr_\futWind$ into account, which leads to a non-causal model estimate as pointed out in~\cite{Shi2001}. Therefore, we follow an alternative route by estimating a predictor in the pre-estimation step and construct  $\obsvTV_\futWind\check\reachTV_\pastWind$, $\check\HankStr_\futWind$, $\tilde\obsvTV_\futWind\tilde\reachTV_\pastWind$ to lower the computational demand and to enforce a causal model, similar to recent literature~\cite{Qin2005,Jansson2005,Chiuso2007a,Verhaegen2007}.

% ############################################### %
% ------------------------------------------------------------------------------ %
% 		Subspace identification: the open-loop data-equation
% ------------------------------------------------------------------------------ %
% ############################################### %
\vspace{-2mm}
\section{Subspace identification in open-loop form} \label{sec:SIDOLTimeInv} \vspace{-2mm}

In this section, we derive two methods to realize the state-sequence based on the open-loop data-equation~\eqref{eq:DataEqOL}. The first method is based on a maximum-likelihood argument using \emph{canonical correlation analysis} (CCA) (Sec.~\ref{subsec:CCA}) and the second method applies a realization based argument (Sec.~\ref{subsec:stochasticReal}). The latter deterministic state realization approach results in the LPV extension of various LTI schemes by using different weighting matrices in the state realization step.

%Both schemes make us of the fact that the observability and reachability matrices can be decomposed in a parameter independent and parameter dependent structure due to the affine parameter structure of~\eqref{eq:SSrepInnAff}. These insights are also used in Sections~\ref{sec:SIDCLTimeInv} and~\ref{sec:BasisReduced} to realize the state-sequence in the closed-loop case.

% ####################################################
% Subsec: Main concept
% ####################################################
\vspace{-2mm}
\subsection{Main concept}  \label{subsec:SIDOLTimeInvMain} \vspace{-2mm}

The stochastic and the deterministic approaches use the fact that the observability and reachability matrices can be decomposed into a parameter independent and a parameter dependent part. To this end, define \vspace{-3mm}
\begin{align*}
\check P^u_{t\vert \pastWind} &= \psi_t \otimes \ldots \otimes \psi_{t-\pastWind} \otimes \eye_\NU, \\
\check P^\xi_{t\vert \pastWind} &=\psi_t \otimes \ldots \otimes \psi_{t-\pastWind} \otimes \eye_{\NY}, \\
\check M_{t,\pastWind} &  =\diag\big(  \check P^u_{t-1\vert0},\check P^\xi_{t-1\vert0},  \ldots, \check P^u_{t-1\vert \pastWind-1},\check P^\xi_{t-1\vert \pastWind-1}\big), \\
L_{t\vert \futWind} &=  \psi^\top_t \otimes \ldots \otimes \psi^\top_{t+\futWind}\otimes \eye_{\NY}, \\
 N_{t,\futWind} &= \diag\left( L_{t\vert 0},\ldots,L_{t\vert \futWind-1}  \right).
\end{align*} \vskip -7mm \noindent
The $\pastWind$-step extended reachability matrix and the $\futWind$-step extended observability matrix  are given as \vspace{-4mm}
\begin{equation} \label{eq:ObsvReach}
\reach_\pastWind = \left[ \begin{array}{ccc} \M_1&\cdots&\M_j  \end{array}   \right], \hspace{0.7cm} \obsv_\futWind = \left[ \begin{array}{ccc} \N_1^\top&\cdots&\N_i^\top  \end{array}   \right]^\top,
\end{equation}\vskip -6mm \noindent
with dimensions $\reach_\pastWind\in\mathbb{R}^{\NX\times\left(\NU\sum_{l=1}^\pastWind(1+\NPSI)^l \right)}$ and $\obsv_\futWind\in\mathbb{R}^{\left( \NY \sum_{l=1}^\futWind(1+\NPSI)^l \right)\times\NX}$ where $\M_k$, $\N_k$ are defined as \vspace{-3mm}
\begin{subequations}  \label{ch8-eq:ReachM}
\begin{align}
  \M_1 &=\left[ \begin{array}{cccccc} B_0&\cdots&B_\NPSI & K_0&\cdots&K_\NPSI  \end{array}   \right],\nonumber\\
  \M_k &=\left[ \begin{array}{ccc} A_0\M_{k-1}&\cdots&A_\NPSI\M_{k-1}  \end{array}   \right], \\
  \N_1 &=\left[ \begin{array}{ccc} C_0^\top&
 \cdots&C_\NPSI^\top  \end{array}\   \right]^\top\!\!, \nonumber \\
  \N_k &=\left[ \begin{array}{ccc} (\N_{k-1}A_0)^\top&\!\!\cdots\!\!&(\N_{k-1}A_\NPSI)^\top  \end{array}   \right]^\top \!\!.
\end{align} \vskip -9mm \noindent
\end{subequations}

Using Assumption~\ref{ass:initialCond}, the open-loop data-equation~\eqref{eq:DataEqOL} can be decomposed as \vspace{-3mm}
\begin{equation} \label{eq:DataEqAffOL} % be casefull with the vphantom to vertically align the underbrace
\futDat{y}{\futWind}{t} = \underbrace{\vphantom{\check\reach_\pastWind \check M_{t,\pastWind}} N_{t,\futWind}\obsv_\futWind}_{\vphantom{(\check\reachTV_\pastWind \diamond p)_t} (\obsvTV_\futWind \diamond p)_t}\underbrace{\vphantom{N_{t,\futWind}\obsv_\futWind}  \check\reach_\pastWind \check M_{t,\pastWind}}_{\vphantom{(\obsvTV_\futWind \diamond p)_t} (\check\reachTV_\pastWind \diamond p)_t} \pastDatc{z}{t}{\pastWind} + (\check\HankStr_\futWind\diamond p)_t  \futDatc{z}{\futWind}{t} + \futDat{\xi}{\futWind}{t},
\end{equation}\vskip -9mm \noindent

Data-equation~\eqref{eq:DataEqAffOL} describes the IO relations of the data-generating system based on an SS form. The unknowns in this IO relation are the so-called \emph{sub-Markov} parameters $C_iA_j\cdots A_kB_l$ and $C_iA_j\cdots A_kK_l$. Using the relation~\eqref{eq:DataEqAffOL}, the sub-Markov parameters and the unknown noise sequence $\xi_t$ can be estimated by LPV-MAX model estimation using convex optimization~\cite[Thm. 5.5]{Cox2018PHD}.

In this section, the state realization is accomplished by assuming that a sub-part of the structural observability matrix $\obsv_\futWind$ associated with the parameter independent part of the SS representation, i.e., $C_0$ and $A_0$, is full column rank (common assumption applied in practice~\cite{Luspay2009,Wingerden2009,Schulz2016,Verdult2002c}).%
\footnote{Any $C_iA_iK_i$ combination could be taken instead of $C_0A_0K_0$. In such case, additional assumptions should be taken on the associated scheduling variable to fulfill the observability criterion, which is not treated to simplify the discussion.} %
To this end, define the scheduling independent observability matrices \vspace{-2mm}
\begin{equation*}
		\obsv^0_\futWind \!=\! \left[ \!\begin{array}{c}  C_0 \\ C_0A_0 \\ \vdots \\ C_0A^{\futWind-1}_0 \end{array} \!  \right], \hspace{0.2cm} \tilde\obsv^0_\futWind\!=\!\left[ \!\begin{array}{c}  C_0 \\ C_0(A_0-K_0C_0) \\ \vdots \\ C_0(A_0-K_0C_0)^{\futWind-1} \end{array} \!  \right],
\end{equation*}\vskip -5mm \noindent
for the open-loop and closed-loop setting, respectively. %
\footnote{The closed-loop scheduling independent observability matrix is presented here for compactness of the paper.}
\vspace{-2mm}
\begin{assumption}
	\item\label{ass:FullObsTimeIndp} The scheduling independent part of the observability matrix is of rank $\NX$, i.e., $\rank(\obsv^0_\futWind)=\NX$ or $\rank(\tilde\obsv^0_\futWind)=\NX$ for the open-loop or closed-loop case, respectively.	
\end{assumption}\vskip -2mm \noindent
%
%
%Assumption~\ref{ass:FullObsTimeIndp} implies that the state is fully observable by the time-invariant part, which is a common assumption in LPV subspace identification and has been successfully applied in some practical applications~\cite{Luspay2009,Wingerden2009,Schulz2016,Kulcsar2009,Verdult2002c}. 
%
%Assumption~\ref{ass:FullObsTimeIndp} will be relaxed in Section~\ref{sec:BasisReduced}.
%
To make use of this assumption,  the observability matrix $\obsv_\futWind$ in~\eqref{eq:DataEqAffOL} is split into a part that depends on $\obsv^0_\futWind$ and a part that does not: \vspace{-4mm}
\begin{multline} \label{eq:DataEqAffOLTIObsv}
\futDat{y}{\futWind}{t}  - (\check\HankStr_\futWind \diamond p)_t \futDatc{z}{\futWind}{t} - N^\ast_{t,\futWind}\obsv^\ast_\futWind\check\reach_\pastWind \check M_{t,\pastWind}\pastDatc{z}{t}{\pastWind} \\
= \obsv^0_\futWind\check\reach_\pastWind \check M_{t,\pastWind}\pastDatc{z}{t}{\pastWind} +  \futDat{\xi}{\futWind}{t},
\end{multline}\vskip -7mm \noindent
where \vspace{-3mm}
\begin{equation*}
\begin{aligned}
L^\ast_{t\vert i} &\!=\!  \psi^\top_t \! \otimes \ldots \otimes \psi^\top_{t+i-1}\!\otimes \big[ \begin{array}{ccc} \psi^\ind{1}_{t+i} & \cdots&\psi^\ind{\NPSI}_{t+i} \end{array} \big]\!\otimes \eye_{\NY}, \\
N^\ast_{t,i} &\!=\! \diag\left( L^\ast_{t\vert 0},\ldots,L^\ast_{t\vert i-1}  \right), \\
\N^\ast_1 &\!=\!\left[ \!\! \begin{array}{ccc} C_1^\top&\cdots&C_\NPSI^\top  \end{array}  \!\! \right]^\top\!\!\!, \hspace{0.2cm} \N^\ast_i \!=\!\left[ \!\!\begin{array}{ccc} (\N^\ast_{i-1}A_0)^{\!\top}&\!\!\cdots\!\!&(\N^\ast_{i-1}A_\NPSI)^{\!\top}  \end{array} \!\!  \right]^\top\!\!\!, \\
 \obsv^\ast_\futWind &\!=\! \left[ \!\! \begin{array}{ccc} (\N^\ast_1)^\top~\cdots~(\N^\ast_\futWind)^\top  \end{array}  \!\! \right]^\top\!\!\!.
\end{aligned}
\end{equation*}\vskip -7mm \noindent
Based on~\eqref{eq:DataEqAffOLTIObsv}, introduce the data-equation describing the so-called \emph{open-loop corrected future}: \vspace{-3mm}
\begin{equation} \label{eq:corrFutOL}
\futDatcCor{y}{\futWind}{t} = \futDat{y}{\futWind}{t}  - (\check\HankStr_\futWind \diamond p)_t \futDatc{z}{\futWind}{t} - N^\ast_{t,\futWind}\obsv^\ast_\futWind\check\reach_\pastWind \check M_{t,\pastWind}\pastDatc{z}{t}{\pastWind}.
\end{equation}\vskip -5mm \noindent
Using an LPV-MAX estimate of~\eqref{eq:DataEqAffOL}, the open-loop corrected future $\futDatcCor{y}{\futWind}{t}$~\eqref{eq:corrFutOL} can be efficiently computed from data. Then, using this surrogate variable, \eqref{eq:DataEqAffOLTIObsv} can be reduced to a data-equation excluding the time-variation in the observability matrix: \vspace{-3mm}
\begin{equation}  \label{eq:DataEqAffOLTIObsvWithCorrFut}
\futDatcCor{y}{\futWind}{t} = \obsv^0_\futWind\check\reach_\pastWind \check M_{t,\pastWind}\pastDatc{z}{t}{\pastWind} +  \futDat{\xi}{\futWind}{t}.
\end{equation}\vskip -5mm \noindent
Representation~\eqref{eq:DataEqAffOLTIObsvWithCorrFut} forms the starting point for finding an estimate of the state-bases. In the sequel, we formulate projection SID methods based on a maximum-likelihood argument (Sec.~\ref{subsec:CCA}) and realization based argument (Sec.~\ref{subsec:stochasticReal}) on the corrected open-loop data-equation~\eqref{eq:DataEqAffOLTIObsvWithCorrFut} to obtain an estimate of the state-sequence.

% ####################################################
% Subsec: Maximum-likelihood estimation
% ####################################################
\vspace{-2mm}
\subsection{Maximum-likelihood estimation} \label{subsec:CCA} \vspace{-2mm}

The corrected formulation~\eqref{eq:DataEqAffOLTIObsvWithCorrFut} is the fundamental data-equation to obtain an estimate of the state-sequence. In this section, the state-sequence is estimated using the canonical correlation analysis. The CCA is a well-known method in statistics that finds a (lower dimensional) space that maximizes the correlation between two random variables~\cite{Roa1979}. In our case, this translates to the objective of finding the unknowns $\obsv^0_\futWind$, $\check\reach_\pastWind$, and the subspace of $x_t$ by maximizing the correlation between $\futDatcCor{y}{\futWind}{t}$ and $\check M_{t,\pastWind}\pastDatc{z}{t}{\pastWind}$, e.g., see~\cite{Chiuso2007a,Gicans2009,Katayama2005,Larimore1983,Larimore2005} to mention a few. In~\cite{Chiuso2007a,Katayama2005} statistical optimality of CCA in the LTI setting has been shown by formulating the optimal one-step-ahead predictor of the state based on either the past or future data. We will take an alternative viewpoint by formulating an estimate of the state-sequence by maximizing the log-likelihood function associated with the \emph{least-squares} (LS) estimation problem of the unknowns $\obsv^0_\futWind\check\reach_\pastWind$ based on the signals $\futDatcCor{y}{\futWind}{t}$ and $\check M_{t,\pastWind}\pastDatc{z}{t}{\pastWind}$ of the model~\eqref{eq:DataEqAffOLTIObsvWithCorrFut}. \cite{Larimore1983,Larimore2005} claim maximum log-likelihood of the state estimation using CCA, however, the mathematical derivations are scattered within the literature and appear to be incomplete, as pointed out in~\cite{Gicans2009}. In Theorem~\ref{lem-eq:CCAnew}, we prove the maximum log-likelihood property for the LPV case. For notational simplicity, let us define the following data-matrices \vspace{-4mm}
\begin{align*}
\check Z_{\pastWind,N} &=\left[\! \begin{array}{ccc} \check M_{1,\pastWind}\pastDatc{z}{1}{\pastWind} & \cdots & \check M_{N,\pastWind}\pastDatc{z}{N}{\pastWind} \end{array}\! \right],\\
\check Y_{\futWind,N}^{(\mathrm{c})}&=\left[ \!\begin{array}{ccc} \futDatcCor{y}{\futWind}{1} & \cdots & \futDatcCor{y}{\futWind}{N} \end{array} \!\right].
\end{align*}\vskip -5mm \noindent

\begin{thm}[CCA based state estimation: open-loop case] \label{lem:LPV-CVA}
Given an LPV data-generating system~\eqref{eq:SSrep} and an associated data set $\Dat$ with $\check Z_{\pastWind,N} \check Z_{\pastWind,N}^\top\succ0$. Compute the following \emph{singular value decomposition} (SVD) \vspace{-4mm}
\begin{multline} \label{lem-eq:CCAnew}
\Bigl(\frac{1}{N} \check Y_{\futWind,N}^{(\mathrm{c})}\bigl(\check Y_{\futWind,N}^{(\mathrm{c})}\bigr)^\top\Bigr)^{-\frac{1}{2}} \check Y_{\futWind,N}^{(\mathrm{c})}  \check Z_{\pastWind,N}^\top \Bigl(\frac{1}{N} \check Z_{\pastWind,N} \check Z_{\pastWind,N}^\top\Bigr)^{-\frac{1}{2}}     \\ = U\tilde SV^\top,
\end{multline}\vskip -8mm \noindent
with the matrices $\tilde U$ and $\tilde V$, given by \vspace{-3mm}
\begin{equation*} % \label{eq:ConstTildeUV}
\tilde U = \Bigl(\frac{1}{N} \check Y_{\futWind,N}^{(\mathrm{c})}\bigl(\check Y_{\futWind,N}^{(\mathrm{c})}\bigr)^{\!\top}\Bigr)^{-\frac{1}{2}} U, \hspace{0.5cm} \tilde V = \Bigl(\frac{1}{N}\check Z_{\pastWind,N} \check Z_{\pastWind,N}^\top\Bigr)^{-\frac{1}{2}} V.
\end{equation*}\vskip -5mm \noindent
Under Assumptions~\ref{ass:ueuncorrOL}-\ref{ass:peuncorrOL} and~\ref{ass:datInM}-\ref{ass:FullObsTimeIndp},
 \vspace{-3mm}
\begin{equation} \label{lem-eq:CCAstate}
\hat X_N = \tilde V_\NX^\top \check Z_{\pastWind,N}, \hspace{0.5cm} \mbox{with} \hspace{0.5cm} \frac{1}{N} \hat X_N \hat X_N^\top = \eye_\NX,
\end{equation}\vskip -7mm \noindent
where $\tilde V_\NX$ defines the first $\NX$ columns of $\tilde V$, is a maximum-likelihood estimate of the state-sequence. The associated log-likelihood function minimized by this estimate is \vspace{-3mm}
\begin{multline} \label{lem-eq:loglikelihoodCVA}
-\log L =  \frac{\futWind\,\NY N}{2} \left( \log(2\pi) + 1\right) - \frac{N}{2} \log\bigl(\det\bigl(\tilde U\bigr)^2\bigr) \\
+ \frac{N}{2} \sum_{i=1}^\NX  \log(1-\tilde \sing^2_i ),
\end{multline} \vskip -7mm
where $\tilde S=\diag(\tilde \sing_1,\ldots,\tilde \sing_\NX )$.\hfill $\square$
\end{thm}\vspace{-2mm}
\begin{pf}
See the Appendix. \hfill $\blacksquare$
\end{pf} \vspace{-5mm}

In case of infinite data, i.e., $N\rightarrow\infty$, $\tilde S$ will contain exactly $\NX$ nonzero singular values which are equal to one (see \cite[Remark 9.1]{Cox2018PHD}). In case of finite data, the state order $\NX$ can be selected by a gap in magnitude between the singular values~\cite{Overschee1996,Verhaegen2007}. Alternatively, the stochastic interpretation of the CCA in Theorem~\ref{lem:LPV-CVA} allows for a data-driven selection of the model order $\NX$  based on the log-likelihood function $L$ \eqref{lem-eq:loglikelihoodCVA} using an information criterion such as Akaike's or the Bayesian information criterion~\cite{Ljung1999}. In depth investigation of order selection is beyond of the scope of this paper.

%Compared to normal SVD, the matrices $\tilde U$ and $\tilde V$ in Th.~\ref{lem:LPV-CVA} are not unitary. 

Using the estimate of the state-sequence $\hat x$, the state-space matrices $\{A_i,B_i,C_i,D_i,K_i\}_{i=0}^\NPSI$ are estimated using two linear regression steps, e.g., see~\cite[Sec. 2.5]{Verdult2002b}. The first step is a standard $\ell_2$-loss minimization problem based on the output-equation~\eqref{eq:SSrepOut} with the solution: \vspace{-4mm}
\begin{equation} \label{eq:LSoutEq}
\left[\! \begin{array}{cccccc} \hat C_0 & \!\!\cdots\!\! & \hat C_\NPSI & \hat D_0 & \!\!\cdots\!\! & \hat D_\NPSI \end{array}\! \right] = Y_N \Phi^\dagger_{\!N,\mathrm{o}},\end{equation}\vskip -6mm \noindent
where $\Phi^\dagger$ defines the right-pseudo inverse of $\Phi$ and the regression matrices are \vspace{-4mm}
\begin{equation*}
\Phi_{N,\mathrm{o}} = \left[\begin{array}{ccc} \psi_1\otimes \hat x_1 & \!\!\cdots\!\! & \psi_N\otimes \hat x_N \\  \psi_1\otimes u_1 &  \!\!\cdots\!\! & \psi_N\otimes u_N \end{array}  \right]\!,\hspace{0.2cm} Y_N = [\begin{array}{ccc}y_1&  \!\!\cdots\!\! & y_N \end{array}].
\end{equation*}\vskip -6mm \noindent
Using the output-equation~\eqref{eq:SSrepOut}, an estimate of the innovation noise is found in the form of the residual error of~\eqref{eq:LSoutEq}: \vspace{-2mm}
\begin{equation}
[\!\begin{array}{ccc} \hat \xi_1 & \!\!\cdots\!\! & \hat \xi_N \end{array}\!] = Y_N \!-\! \left[\!\! \begin{array}{cccccc} \hat C_0 & \!\cdots\! & \hat C_\NPSI & \hat D_0 & \!\cdots\! & \hat D_\NPSI \end{array}\!\! \right] \! \Phi_{N,\mathrm{o}}.
\end{equation}\vskip -6mm \noindent
The remaining state-space matrices are estimated by a second linear-regression step based on the state equation~\eqref{eq:SSrepState}: \vspace{-3mm}
\begin{equation} \label{eq:LSstateEq}
\left[\!\! \begin{array}{ccccccccc} \hat A_0 & \!\!\cdots\!\! & \hat A_\NPSI & \!\!\hat B_0 & \!\!\cdots\!\! & \hat B_\NPSI & \!\!\hat K_0 & \!\!\cdots\!\! & \hat K_\NPSI \end{array}\!\! \right] = \hat X^\prime_N \Phi^\dagger_{\!N,\mathrm{s}},
\end{equation}\vskip -5mm \noindent
with \vspace{-3mm}
\begin{equation*}
\Phi_{N,\mathrm{s}} \!=\! \left[\!\begin{array}{ccc} \psi_1\otimes \hat x_1 & \!\!\cdots\!\! & \psi_{N-1}\otimes \hat x_{N-1} \\ \psi_1\otimes u_1 & \!\!\cdots\!\! & \psi_{N-1}\otimes u_{N-1} \\ \psi_1\otimes \hat \xi_1 & \!\!\cdots\!\! & \psi_{N-1}\otimes \hat \xi_{N-1} \end{array}\!  \right]\!, \hspace{0.2cm} \hat X^\prime_N \!=\! \left[\!\begin{array}{ccc}\hat x_2 & \!\!\cdots\!\! &  \hat x_N \end{array} \!  \right]\!.
\end{equation*}\vskip -7mm \noindent

% ####################################################
% Subsec: Realization based estimation
% ####################################################
\vspace{-2mm}
\subsection{Realization based estimation} \label{subsec:stochasticReal} \vspace{-2mm}

In Section~\ref{subsec:CCA}, a statistical viewpoint has been taken based on only the input-scheduling-corrected output signals. Alternatively, inspired by the Ho-Kalman scheme in~\cite{Toth2012} or the PBSID scheme in~\cite{Wingerden2009}, the problem can be tackled from the realization point of view, i.e., the state is realized by decomposing the sub-Markov coefficients in $\obsv^0_\futWind\check\reach_\pastWind$. Opposed to the Ho-Kalman scheme, we are interested in the state-sequence of the innovation form~\eqref{eq:SSrep} including noise dynamics. In the LTI case, these ideas have been extensively exploited resulting in many variants of subspace identification schemes, e.g., see \cite{Overschee1996,Qin2006,Verhaegen2007}. As recognized in \cite[Thm. 12]{Overschee1996}, most of the LTI subspace schemes only differ by left- and right-multiplication of the Hankel matrix with different weightings. Following this concept, a unified LPV formulation of subspace schemes can be introduced:

\begin{thm}[Unified state realization: the open-loop case] \label{lem:UnifiedOL}
Given an LPV data-generating system~\eqref{eq:SSrep} and an associated data set $\Dat$ with $\check Z_{\pastWind,N} \check Z_{\pastWind,N}^\top\succ0$. Under Assumptions~\ref{ass:ueuncorrOL}-\ref{ass:peuncorrOL} and~\ref{ass:datInM}-\ref{ass:FullObsTimeIndp}, let $\obsv^0_\futWind\check\reach_\pastWind$ be the consistent estimate of the sub-Hankel matrix in~\eqref{eq:DataEqAffOLTIObsvWithCorrFut}. Compute the following SVD \vspace{-2mm}
\begin{equation}
W_1 \obsv^0_\futWind\check\reach_\pastWind W_2 = USV^\top,
\end{equation}\vskip -6mm \noindent
where the full rank weightings can be taken as \vspace{-4mm}
\begin{align*}
\mbox{{HK} } &\left\{ \begin{array}{l} W_1 \!=\! \eye, \\ W_2 \!=\! \eye,  \end{array} \right. \hspace{1cm}
\mbox{N4SID }\!\left\{ \!\!\begin{array}{l} W_1 \!=\! \eye, \\ W_2 \!=\! \left(\check Z_{\pastWind,N}\check Z_{\pastWind,N}^\top\right)^{\frac{1}{2}}\!\!\!\!,  \end{array} \right. \\
\mbox{p-CCA }&\left\{ \begin{array}{l} W_1 \!=\! \Big(\check Y_{\futWind,N}^{(\mathrm{c})}\left(\check Y_{\futWind,N}^{(\mathrm{c})}\right)^{\!\!\!\top} \Big)^{\!\!-\frac{1}{2}}\!\!\!\!, \\[2mm] W_2 \!=\! \left(\check Z_{\pastWind,N}\check Z_{\pastWind,N}^\top\right)^{\!\!\frac{1}{2}}\!\!.  \end{array} \right.
\end{align*}\vskip -6mm \noindent
The \emph{Ho-Kalman} (HK), \emph{numerical subspace state space system identification} (N4SID) and \emph{projected canonical correlation analysis} (p-CCA) follow the default naming in subspace literature, see~\cite{Overschee1996}. Then, a realization of the state-sequence is given by \vspace{-1mm}
\begin{equation} \label{lem-eq:UnistateOL}
\hat X_N = S^{\frac{1}{2}}V_\NX^\top W_2^{-1} \check Z_{\pastWind,N}.
\end{equation}\vskip -3mm \noindent
\hfill $\square$
\end{thm}\vspace{-5mm}
\begin{pf}
The LPV estimation problem can be rewritten in an LTI formulation, because the IO model~\eqref{eq:DataEqAffOLTIObsvWithCorrFut} representing the SS form of the data-generating system is linear time-invariant w.r.t. the signals $\check M_{t,\pastWind}\pastDatc{z}{t}{\pastWind}$ and $\futDatcCor{y}{\futWind}{t}$. Hence, the derivation for LTI subspace schemes can be directly applied. For a rigorous proof, see \cite[Chpt. 4.3]{Overschee1996}. To illustrate, it is not difficult to show that \vspace{-2mm}
\begin{equation} \label{lem-pf-eq:realizationObsvReach}
\obsv^0_\futWind = W_1^{-1}U_\NX S^{\frac{1}{2}}_\NX, \hspace{1cm}\check\reach_\pastWind = S^{\frac{1}{2}}_\NX V^\top_\NX W_2^{-1}.
\end{equation}\vskip -5mm \noindent
Hence, taking~\eqref{eq:StateOL} and Assumption~\ref{ass:initialCond} into account, right-multiplying the reachability matrix $\check\reach_\pastWind$ in~\eqref{lem-pf-eq:realizationObsvReach} with the data matrix $\check Z_{\pastWind,N}$ leads to a realization of the state as in~\eqref{lem-eq:UnistateOL}. \hfill $\blacksquare$
\end{pf} \vspace{-5mm}

An important fact in Theorem~\ref{lem:UnifiedOL} is the absence of the closed-loop dynamics, contrary to the LTI case~\cite{Overschee1996,Qin2006}, and the required pre-estimation step to obtain  $\obsv^0_\futWind\check\reach_\pastWind$. Opposed to the LTI case, we do not apply oblique projections to remove the effect of future inputs, e.g., see~\cite[Sec. 4.2]{Overschee1996} (the oblique projection is an indirect LS estimation and prediction step). In the LPV case, we apply the pre-estimation step making the oblique projections of the future input superfluous and, therefore, a MOESP like weighting is not present. The unified formulation in Theorem~\ref{lem:UnifiedOL} applies N4SID and CCA like weightings to the estimated Hankel matrix, but it is not an LPV extension of these methods, due to the missing oblique projections. In addition, it is important to note that the CCA in Theorem~\ref{lem:LPV-CVA} and the p-CCA in Theorem~\ref{lem:UnifiedOL} are different as CCA is based on stochastic realization theory and signal relations while p-CCA is based on pre-estimated sub-Markov coefficients. This theoretical split can also be found in the LTI literature, e.g., between~\cite{Larimore1990b} and \cite{Overschee1996}, respectively. Both principles are equivalent for $N\rightarrow\infty$, as the oblique projections and least-squares estimates are consistent and unbiased~\cite{Bauer2002}.  The choice of the weightings $W_1$ and $W_2$ has been discussed by many authors. In the LTI case, it has been proven that  $W_1$ has no influence on the asymptotic accuracy of the estimates, see~\cite{Bauer2002,Gustafsson2002,Chiuso2004}. On the other hand, on finite data, the optimal choice is still an open question.

For any applied weighting, the estimated state-sequence in the unified formulation \eqref{lem-eq:UnistateOL} is not guaranteed to have unit variance compared to the estimate by the CCA method in \eqref{lem-eq:CCAstate}. In the LTI case, it can be shown that the resulting model estimate is stochastically balanced for any choice of the weighting~\cite{Arun1990}, similar to deterministic Ho-Kalman realization. In the LPV case, the observability and reachability Gramians are scheduling-dependent and the authors believe that the state-sequence~\eqref{lem-eq:UnistateOL} is structurally balanced, but formally proving this property is a subject of future research.

% ############################################### %
% ------------------------------------------------------------------------------ %
% 		Subspace identification: the closed-loop data-equation
% ------------------------------------------------------------------------------ %
% ############################################### %

\section{Subspace identification in closed-loop form} \label{sec:SIDCLTimeInv}\vspace{-2mm}

The concepts of the presented state estimation and realization schemes for the open-loop identification setting in Th.~\ref{lem:LPV-CVA} and~\ref{lem:UnifiedOL} will be extended to the closed-loop case in this section. Similar to the open-loop case, the realization problem is first tackled from an ML point of view (Sec.~\ref{subsec:CCACL}) and then from a deterministic realization viewpoint (Sec.~\ref{subsec:stochasticRealCL}). We would like to emphasize that the scheme presented in~\cite{Wingerden2009a} simplifies the realization problem by considering the matrix functions $C$, $D$, and $K$ constant. No such assumption will be taken next.

% ############################################### %
% 		Main concept
% ############################################### %
\vspace{-2mm}
\subsection{Main concept} \label{subsec:SIDCLTimeInvMain} \vspace{-2mm}

Construction of the matrices $\tilde N_{t,\futWind}$, $\tilde\obsv_\futWind$, $\tilde\reach_\pastWind$, and $\tilde M_{t,\pastWind}$ in the closed-loop case are more involved due to the multi-quadratic parameterization of $\tilde\Afnc(p_t)$ and $\tilde\Bfnc(p_t)$ in~\eqref{eq:SSMatStateCL}. First, define all unique combinations of the scheduling induced variation $\psi_t\otimes\psi_t$ as \vspace{-3mm}
\begin{multline} \label{eq:extScheduling}
\mu_t \!=\! \Bigl[\! \begin{array}{cccccc} 1& \psi^\top_t\! & \psi^{\ind{1}}_t \psi^{\ind{1}}_t & \!\!\cdots\!\! & \psi^{\ind{1}}_t \psi^{\ind{\NPSI}}_t\!  & \psi^{\ind{2}}_t \psi^{\ind{2}}_t\! \end{array} \\
\begin{array}{ccc} \psi^{\ind{2}}_t \psi^{\ind{3}}_t\! & \!\!\cdots\!\! & \psi^{\ind{\NPSI}}_t \psi^{\ind{\NPSI}}_t \end{array} \! \Bigr]^\top\!\!\!\!,
\end{multline} \vskip -7mm \noindent
where $\mu:\mathbb{Z}\rightarrow\mathbb{M}\subset\mathbb{R}^{\NMU+1}$ with dimension $\NMU=\frac{1}{2}\NPSI(\NPSI+3)$ is called the \emph{extended scheduling variable}%
\footnote{$\NMU$ is given by {\tiny$\left( \begin{array}{c} \NPSI+1 \\ 1 \end{array} \right)+\left( \begin{array}{c} \NPSI+1 \\ 2 \end{array} \right)-1$} where {\tiny$\left( \begin{array}{c} n \\ k \end{array} \right)$} denotes the binomial coefficient.} in the sequel. %
In addition, define \vspace{-3mm}
\begin{multline*}
 \big[\begin{array}{ccc}\ACoef_{0} & \cdots & \ACoef_{\NMU} \end{array}\big]\! = \!
\big[\begin{array}{ccc} A_0\!-\!K_0C_0& ~~A_1\!-\!K_1C_0\!-\!K_0C_1 &  \cdots \end{array} \\ 
\begin{array}{cccc} A_\NPSI\!-\!K_\NPSI C_0\!-\!K_0C_\NPSI & ~~-K_1C_1 & ~~-K_1C_2\!-\!K_2C_1 & \cdots \end{array} \\ 
\begin{array}{cccc} -K_1C_\NPSI\!-\!K_\NPSI C_1 & ~~-K_2C_2 & ~~-K_2C_3\!-\!K_3C_2 &\cdots  \end{array}\big],
\end{multline*}\vskip -6mm \noindent
and \vspace{-2mm}
\begin{multline*}
 \big[\begin{array}{ccc}\BCoef_{0} & \cdots & \BCoef_{\NMU} \end{array}\big]\! =\! 
\big[\begin{array}{ccc} B_0\!-\!K_0D_0& ~~B_1\!-\!K_1D_0\!-\!K_0D_1 &  \cdots \end{array} \\ 
\begin{array}{cccc} B_\NPSI\!-\!K_\NPSI D_0\!-\!K_0D_\NPSI & ~~-K_1D_1 & ~~-K_1D_2\!-\!K_2D_1 & \cdots \end{array} \\ 
\begin{array}{cccc} -K_1D_\NPSI\!-\!K_\NPSI D_1 & ~~-K_2D_2 & ~~-K_2D_3\!-\!K_3D_2 &\cdots  \end{array}\big],
\end{multline*}\vskip -7mm \noindent
Next, let us define the closed-loop $\pastWind$-step extended reachability matrix: \vspace{-3mm}
\begin{subequations} \label{eq:ObsvReach-CL}
\begin{align}
\tilde\M^u_1=& \left[ \begin{array}{ccc} \BCoef_0  & \cdots & \BCoef_{\NMU} \end{array} \right], \hspace{1cm} \tilde\M^y_1= \left[ \begin{array}{ccc} K_0&\cdots&K_\NPSI \end{array} \right], \nonumber  \\
\tilde\M^{\mathpzc{i}}_j=&\left[\!\! \begin{array}{ccc} \ACoef_0 & \!\cdots\! &  \ACoef_{\NMU} \end{array} \!\!\right] ( \eye_\NMU\otimes\tilde\M^{\mathpzc{i}}_{j-1}), \nonumber\\
\tilde\reach_\pastWind =& \left[ \begin{array}{ccccc} \tilde\M^u_1 & \tilde\M^y_1 & \cdots & \tilde\M^u_\pastWind & \tilde\M^y_\pastWind \end{array} \right],
\end{align}\vskip -5mm \noindent
with $\mathpzc{i}\in\{u,y\}$ and define the closed-loop $\futWind$-step extended observability matrix as \vspace{-4mm}
\begin{align}
\tilde\N_1 &\!=\!\left[ \!\! \begin{array}{ccc} C_0^\top&\!\cdots\!&C_\NPSI^\top  \end{array}  \!\! \right]^\top\!\!\!\!, \hspace{0.5cm} \tilde\N_j \!= (\eye_\NMU\!\otimes\! \tilde \N_{j-1}) \left[\!\! \begin{array}{ccc} \ACoef^\top_0 & \!\cdots\! &  \ACoef^\top_{\NMU} \end{array} \!\! \right]^\top\!\!\!\!, \nonumber \\
\tilde\obsv_\futWind &\!=\! \left[ \!\! \begin{array}{ccc} \tilde\N_1^\top~\cdots~\tilde\N_\futWind^\top  \end{array}  \!\! \right]^\top\!\!\!\!.
\end{align}\vskip -5mm \noindent
\end{subequations}
Finally, the scheduling dependent data-matrices are given as \vspace{-2mm}
\begin{align*}{}
\tilde P^u_{t\vert j} &= \mu_t \otimes \cdots \otimes \mu_{t-j} \otimes \eye_\NU, \\
\tilde P^y_{t\vert j} &= \mu_t \otimes \cdots \otimes \mu_{t-j-1} \otimes \psi_{t-j} \otimes \eye_{\NY},  \\
\tilde M_{t,j} &  =\diag\big(  \tilde P^u_{t-1\vert0},\tilde P^y_{t-1\vert0},  \cdots, \tilde P^u_{t-1\vert j-1},\tilde P^y_{t-1\vert j-1}\big), \\
\tilde L_{t\vert i}&=  \mu^\top_t \otimes \cdots \otimes \mu^\top_{t+i-1}\otimes \psi^\top_{t+i} \otimes \eye_{\NY}, \\
\tilde N_{t,i} &= \diag\big( \tilde L_{t\vert 0},\cdots, \tilde L_{t\vert i-1}  \big).
\end{align*}\vskip -5mm \noindent
By applying the aforementioned matrices and Assumption~\ref{ass:initialCond}, Eq.~\eqref{eq:DataEqCL} reads as \vspace{-3mm}
\begin{equation} \label{eq:DataEqAffCL} % Be casefull \vphantom makes this equation horrible
\futDat{y}{\futWind}{t} = \underbrace{\vphantom{\tilde\reach_\pastWind \tilde M_{t,\pastWind}} \tilde N_{t,\futWind}\tilde\obsv_\futWind}_{\vphantom{(\tilde\reachTV_\pastWind\diamond p)_t} (\tilde\obsvTV_\futWind\diamond p)_t} \underbrace{\vphantom{\tilde N_{t,\futWind}\tilde\obsv_\futWind}\tilde\reach_\pastWind \tilde M_{t,\pastWind}}_{\vphantom{(\tilde\obsvTV_\futWind\diamond p)_t} (\tilde\reachTV_\pastWind\diamond p)_t} \pastDatt{z}{t}{\pastWind} + (\tilde\HankStr_\futWind \diamond p)_t  \futDatt{z}{\futWind}{t} + \futDat{\xi}{\futWind}{t}.
\end{equation}\vskip -9mm \noindent

It can be proven that the LPV-SS representation~\eqref{eq:SSrep} with state dimension $\NX$ is \emph{stochastically structurally state-observable} if and only if $\rank(\tilde\obsv_\NX)=\NX$~\cite[Lem. 9.4]{Cox2018PHD}. Similarly, the SS representation~\eqref{eq:SSrep} is \emph{stochastically structurally state-reachable} if and only if $\rank(\tilde\reach_\NX)=\NX$~\cite[Lem. 9.4]{Cox2018PHD}. In addition, based on the extended Hankel matrix $\tilde\obsv_\futWind\tilde\reach_\pastWind$, the existence of a stochastic realization of $\sys_\mathrm{o}$ with finite model order $\NX$ can be proven~\cite[Lem. 9.5]{Cox2018PHD}. Hence, the data-equation~\eqref{eq:DataEqAffCL} allows us to obtain a state-space realization of the data-generating system $\sys_\mathrm{o}$ by representation~\eqref{eq:SSrep} . 
%To consistently estimate the parameters in $\tilde\obsv_\futWind\tilde\reach_\pastWind$, we require $\rank(\Xi^2)=\NY$ (persistency of excitation on $\xi_t$) and persistency of excitation on $\tilde M_{t,\pastWind}\pastDatt{z}{t}{\pastWind}$. 
The unknown coefficients $C_i\ACoef_0\cdots\ACoef_0\BCoef_0$, $C_i\ACoef_{ij}\cdots\ACoef_{ij}\BCoef_{ij}$, $C_i\ACoef_0\cdots\ACoef_0K_i$, $C_i\ACoef_{ij}\cdots\ACoef_{ij}K_i$ for $i,j=0,\ldots,\NPSI$ found in $\tilde\obsv_\futWind\tilde\reach_\pastWind$ and $\tilde\HankStr_\futWind$ are the sub-Markov coefficients of the multi-quadratic parameterization of the closed-loop formulation~\eqref{eq:DataEqAffCL}. These unknown quantities can be estimated by a linear regression of an LPV-ARX model.

The developed concepts of the open-loop setting can be applied to obtain a realization of the model in the closed-loop setting. This concept has successfully been used in the LPV literature, e.g., in \cite{Verdult2002b,Wingerden2009a,Larimore2005} to mention a few. To this end, the closed-loop counterpart of~\eqref{eq:DataEqAffOLTIObsv} is \vspace{-4mm}
\begin{multline} \label{eq:DataEqAffCLTIObsv}
\futDat{y}{\futWind}{t}  - (\tilde\HankStr_\futWind\diamond p)_t \futDatt{z}{\futWind}{t} - \tilde N^\ast_{t,\futWind}\tilde\obsv^\ast_\futWind\tilde\reach_\pastWind \tilde M_{t,\pastWind}\pastDatt{z}{t}{\pastWind} \\
= \tilde\obsv^0_\futWind\tilde\reach_\pastWind \tilde M_{t,\pastWind}\futDatt{z}{t}{\pastWind} +  \futDat{\xi}{\futWind}{t},
\end{multline}\vskip -7mm \noindent
where \vspace{-4mm}
\begin{align*}
\tilde\N^\ast_1 &\!=\!\left[ \!\! \begin{array}{ccc} C_1^\top&\!\cdots\!&C_\NPSI^\top  \end{array}  \!\! \right]^\top\!\!\!\!, \hspace{0.4cm}
\tilde\N^\ast_j \!= (\eye_\NMU\otimes \tilde \N^\ast_{j-1}) \left[\!\! \begin{array}{ccc} \ACoef^\top_0 & \!\cdots\! &  \ACoef^\top_{\NPSI} \end{array} \!\! \right]^\top\!\!\!\!, \\
\tilde\obsv^\ast_\futWind &\!=\! \left[ \!\! \begin{array}{ccc} (\tilde\N_1^\ast)^\top~\cdots~(\tilde\N_\futWind^\ast)^\top  \end{array}  \!\! \right]^\top\!\!\!\!, \\
\tilde L^\ast_{t\vert i} &=  \mu^\top_t \otimes \cdots \otimes \mu^\top_{t+i-1}\otimes \big[ \begin{array}{ccc} \psi^\ind{1}_{t+i} & \cdots&\psi^\ind{\NPSI}_{t+i} \end{array} \big] \otimes \eye_{\NY},\\ 
\tilde N^\ast_{t,i} &= \diag\big( \tilde L^\ast_{t\vert 0},\ldots, \tilde L^\ast_{t\vert i-1}  \big).
\end{align*}\vskip -6mm \noindent
From~\eqref{eq:DataEqAffCLTIObsv}, the \emph{closed-loop corrected future} can be introduced as \vspace{-2mm}
\begin{equation}
\futDattCor{y}{\futWind}{t} = \futDat{y}{\futWind}{t} - (\tilde\HankStr_\futWind \diamond p)_t  \futDatt{z}{\futWind}{t} - \tilde N^\ast_{t,\futWind}\tilde\obsv^\ast_\futWind\tilde\reach_\pastWind \tilde M_{t,\pastWind}\pastDatt{z}{t}{\pastWind},
\end{equation}\vskip -5mm \noindent
then using the same principle as in the open-loop case, \eqref{eq:DataEqAffCLTIObsv} is reduced to a data-equation where the time-variation in the observability matrix disappears due to the use of the corrected future \vspace{-2mm}
\begin{equation}  \label{eq:DataEqAffCLTIObsvWithCorrFut}
\futDattCor{y}{\futWind}{t} = \tilde\obsv^0_\futWind\tilde\reach_\pastWind \tilde M_{t,\pastWind}\pastDatt{z}{t}{\pastWind} +  \futDat{\xi}{\futWind}{t}.
\end{equation}\vskip -6mm \noindent
The closed-loop form~\eqref{eq:DataEqAffCLTIObsvWithCorrFut} enables to treat the state realization problem equivalent to the open-loop case in Section~\ref{sec:SIDOLTimeInv}.

% ############################################### %
% 		Maximum-likelihood estimation
% ############################################### %
\vspace{-2mm}
\subsection{Maximum-likelihood estimation} \label{subsec:CCACL} \vspace{-2mm}

The corrected formulation~\eqref{eq:DataEqAffCLTIObsvWithCorrFut} is the fundamental data-equation to obtain an estimate of the state-sequence. In this section, the state-sequence is estimated using the ML point of view introduced in Section~\ref{subsec:CCA}. For notational simplicity, let us define the following data-matrices \vspace{-3mm}
\begin{align*}
\tilde Z_{\pastWind,N} &\!=\! \left[\! \begin{array}{ccc} \tilde M_{1,\pastWind}\pastDatt{z}{1}{\pastWind} & \cdots & \tilde M_{N,\pastWind}\pastDatt{z}{N}{\pastWind} \end{array}\! \right], \\
\tilde Y_{\futWind,N}^{(\mathrm{c})} &\!=\! \left[ \!\begin{array}{ccc} \futDattCor{y}{\futWind}{1} & \cdots & \futDattCor{y}{\futWind}{N} \end{array} \!\right].
\end{align*}\vskip -6mm \noindent
For the closed-loop case, we can reformulate Th.~\ref{lem:LPV-CVA} as:

\begin{thm}[CCA based state estimation: closed-loop case] \label{lem:LPV-SSARX}
Given an LPV data-generating system~\eqref{eq:SSrep} and an associated data set $\Dat$ with $\tilde Z_{\pastWind,N} \tilde Z_{\pastWind,N}^\top\succ0$. Compute the SVD \vspace{-3mm}
\begin{multline} \label{lem-eq:CCAnewCL}
\Bigl(\frac{1}{N} \tilde Y_{\futWind,N}^{(\mathrm{c})}\bigl(\tilde Y_{\futWind,N}^{(\mathrm{c})}\bigr)^\top\Bigr)^{-\frac{1}{2}} \tilde Y_{\futWind,N}^{(\mathrm{c})}  \tilde Z_{\pastWind,N}^\top \Bigl(\frac{1}{N} \tilde Z_{\pastWind,N} \tilde Z_{\pastWind,N}^\top\Bigr)^{-\frac{1}{2}}     \\
= U\tilde SV^\top,
\end{multline}\vskip -7mm \noindent
and the matrices $\tilde U$ and $\tilde V$ given by \vspace{-3mm}
\begin{equation*} %\label{eq:ConstTildeUVCL}
\tilde U \!=\! \Bigl(\frac{1}{N} \tilde Y_{\futWind,N}^{(\mathrm{c})}\bigl(\tilde Y_{\futWind,N}^{(\mathrm{c})}\bigr)^{\!\top}\Bigr)^{-\frac{1}{2}} U, \hspace{0.3cm} \tilde V \!=\! \Bigl(\frac{1}{N}\tilde Z_{\pastWind,N} \tilde Z_{\pastWind,N}^\top\Bigr)^{-\frac{1}{2}} V.
\end{equation*}\vskip -5mm \noindent
Under Assumptions~\ref{ass:ueuncorrCL}-\ref{ass:peuncorrCL} and~\ref{ass:datInM}-\ref{ass:FullObsTimeIndp}, \vspace{-3mm}
\begin{equation} \label{lem-eq:CCAstateCL}
\hat X_N = \tilde V_\NX^\top \tilde Z_{\pastWind,N}, \hspace{0.5cm} \mbox{with} \hspace{0.5cm} \frac{1}{N} \hat X_N \hat X_N^\top = \eye_\NX,
\end{equation}\vskip -6mm \noindent
provides a maximum-likelihood estimate of the state-sequence.
The associated log-likelihood function minimized by this estimate is \vspace{-3mm}
\begin{multline} \label{lem-eq:loglikelihoodSSARX}
-\log L =  \frac{\futWind\NY N}{2} \left( \log(2\pi) + 1\right) \\
- \frac{N}{2} \log\bigl(\det\bigl(\tilde U\bigr)^2\bigr)  + \frac{N}{2} \sum_{i=1}^\NX  \log(1-\tilde \sing^2_i ),
\end{multline} \vskip -7mm
where $\tilde S=\diag(\tilde \sing_1,\ldots,\tilde \sing_\NX)$.\hfill $\square$
\end{thm}\vspace{-4mm}
\begin{pf}
Follows the same reasoning as in Theorem~\ref{lem:LPV-CVA} with trivial adaptations. The complete proof is found in~\cite{Cox2018PHD}. \hfill $\blacksquare$
\end{pf} \vspace{-5mm}

Note that, Theorem~\ref{lem:LPV-SSARX} is the LPV counterpart of the LTI SS-ARX scheme presented in~\cite{Jansson2003}. Hence, an important contribution of our framework is the extension and the generalization of the CCA to the LPV setting making it possible to directly extend the SS-ARX scheme. In addition, as a contribution, the derived CCA setting allows to prove the maximum-likelihood property and to obtain the log-likelihood function of the estimate, which have not been formally proven in the LTI case, see in~\cite{Jansson2003}.

% ############################################### %
% 		Maximum-likelihood estimation
% ############################################### %
\vspace{-2mm}
\subsection{Realization based estimation} \label{subsec:stochasticRealCL} \vspace{-2mm}

The state estimation problem has been tackled from the input-scheduling-corrected output statistics point of view in Theorem~\ref{lem:LPV-SSARX}. Alternatively, the state-sequence realization problem can be interpreted as a weighted decomposition of the stochastic, closed-loop Hankel matrix $\tilde\obsv^0_\futWind\tilde\reach_\pastWind$. More specifically, the concepts introduced for the open-loop case in Section~\ref{subsec:stochasticReal} can be directly extended to the closed-loop case leading to a unified theory, which immediately extends various LTI subspace methods to the LPV case:

\begin{thm}[Unified state realization: closed-loop case] \label{lem:UnifiedCL}
Given an LPV data-generating system~\eqref{eq:SSrep} and an associated data set $\Dat$ with $\tilde Z_{\pastWind,N} \tilde Z_{\pastWind,N}^\top\succ0$. Under Assumptions~\ref{ass:ueuncorrCL}-\ref{ass:peuncorrCL} and~\ref{ass:datInM}-\ref{ass:FullObsTimeIndp}, let $\tilde\obsv^0_\futWind\tilde\reach_\pastWind$ be the consistent estimate of the sub-Hankel matrix in~\eqref{eq:DataEqAffCLTIObsvWithCorrFut}. Compute the following SVD \vspace{-2mm}
\begin{equation} \label{lem-eq:SVDUnifiedCL}
W_1 \tilde\obsv^0_\futWind\tilde\reach_\pastWind W_2 = USV^\top,
\end{equation}\vskip -5mm \noindent
where the full rank weightings can be taken as \vspace{-3mm}
\begin{align*}
\mbox{HK } &\left\{\! \begin{array}{l} W_1 \!=\! \eye, \\ W_2 \!=\! \eye,  \end{array} \right. \hspace{0.7cm}
\mbox{PBSID }\left\{ \!\begin{array}{l} W_1 \!=\! \eye , \\ W_2 \!=\! \left(\tilde Z_{\pastWind,N}\tilde Z_{\pastWind,N}^\top\right)^{\frac{1}{2}}\!\!,  \end{array} \right. \\[2mm]
\mbox{p-SS-ARX}&\left\{\! \begin{array}{l} W_1 \!=\! \Big(\tilde Y_{\futWind,N}^{(\mathrm{c})}\left(\tilde Y_{\futWind,N}^{(\mathrm{c})}\right)^{\!\!\!\top} \Big)^{\!\!-\frac{1}{2}}\!\!, \\[2mm] W_2 \!=\! \left(\tilde Z_{\pastWind,N}\tilde Z_{\pastWind,N}^\top\right)^{\!\!\frac{1}{2}}\!\!.  \end{array} \right.
\end{align*}\vskip -5mm \noindent
The HK, \emph{predictor based subspace identification} (PBSID) and \emph{projected space state autoregressive exogenous method} (p-SS-ARX) follow the default naming in subspace literature, see~\cite{Overschee1996,Wingerden2009a,Jansson2003}. %
%\footnote{HK refers to Ho-Kalman, PBSID to predictor based subspace identification, and p-SS-ARX to the projected space state autoregressive exogenous method~\cite{Overschee1996,Wingerden2009a,Jansson2003}.%
%} 
Then, a realization of the state-sequence is given by \vspace{-3mm}
\begin{equation} \label{lem-eq:UnistateCL}
\hat X_N = S^{\frac{1}{2}}V_\NX^\top W_2^{-1} \tilde Z_{\pastWind,N}.
\end{equation}\vskip -6mm \noindent
\hfill $\square$
\end{thm} \vspace{-5mm}
\begin{pf}
Based on a similar argument as for Theorem~\ref{lem:UnifiedOL}. \hfill $\blacksquare$
\end{pf} \vspace{-5mm}

%\begin{rem}
In~\cite{Chiuso2007a,Wingerden2009a} the implementation and derivation of the PBSID method is accomplished differently. Without exploring the theoretical basis, similar to the above given general theory, the authors in~\cite{Chiuso2007a,Wingerden2009a} aimed at realizing a computationally efficient estimator by computing the SVD on $W_1 \tilde\obsv^0_\futWind\tilde\reach_\pastWind\tilde Z_{\pastWind,N}=USV^\top$ and realize the state by $\hat X_N = S^{\frac{1}{2}}V_\NX^\top$. Obviously, this method is equivalent to the above defined PBSID weighting, but it is computationally more efficient as it avoids the square root operation in~\eqref{lem-eq:SVDUnifiedCL}. Additionally,~\cite{Chiuso2007a} proves asymptotic equivalence between LTI PBSID and LTI SS-ARX. Extension of this proof to the LPV case has not been accomplished yet, but it is likely to hold. % \hfill $\square$
A so-called ``optimal'' formulation of Theorem~\ref{lem:UnifiedCL} can also be derived~\cite[Sec. 9.8]{Cox2018PHD} based on the LTI formulation~\cite{Chiuso2007a}. The general idea of~\cite{Chiuso2007a} is to prove that the initial condition $\initCond{\pastWind}$ on the data-equation falls within the variance of the estimator and it can be neglected %due to the finite data estimation problem, 
if the past window $\pastWind$ is chosen large enough. This concept translates to taking the assumption that $[ \prod_{i=1}^\pastWind\! \Afnc(p_{t-i}) ] \!\approx\! 0$ or $[ \prod_{i=1}^\pastWind (\tilde\Afnc\diamond p)_{t-i} ]\approx 0$ for all $p\in\sP^\sZ$ instead of Assumption~\ref{ass:initialCond}. %Concerning Theorem~\ref{lem:UnifiedFull}, this change in Assumption will only effect the construction of the Hankel matrices and not the realization problem~\cite[Sec. 9.8]{Cox2018PHD}. The ``optimal'' reduced state realization will not be presented due to space limitations.

\begin{rem}
Theorems~\ref{lem:LPV-CVA}, \ref{lem:UnifiedOL}, \ref{lem:LPV-SSARX}, and \ref{lem:UnifiedCL} can straightforwardly be modified such that $\Afnc(p_t),\ldots,\Kfnc(p_t)$ are affinely dependent on individual basis functions $\{(\alpha^\ind{i}\diamond p)_t)\}_{i=1}^{n_\alpha},\ldots,\allowbreak\{(\kappa^\ind{j}\diamond p)_t)\}_{j=1}^{n_\kappa}$ with dynamic dependency. % instead of $\{\psi^\ind{i}(p_t)\}_{i=1}^\NPSI$. %However, we used common basis function for notational simplicity.
\end{rem}

\begin{rem}
To lower the computational load w.r.t. the IO estimation and realization, we can apply the kernelization based computation similar to~\cite{Verdult2005,Wingerden2009a}. %The computational gain is obtained when we only use the product between the sub-Markov coefficients and data matrix. For the CCA case, we can compute the corrected future using the dual formulation and, subsequently, kernel canonical correlation analysis. For the realization based case, it would imply a slight modification of Lemma~\ref{lem:UnifiedFull} where: $\hank_{\selO,\selR} = \obsv_\selO\check\reach_\selR S_\selR \check Z_{\pastWind,N}$ or $\hank_{\selO,\selR} = \tilde\obsv_\selO\tilde\reach_\selR S_\selR \tilde Z_{\pastWind,N}$. In this case, we are selecting only specific rows and columns out of the full matrix $\obsv_\futWind\tilde\reach_\pastWind\tilde Z_{\pastWind,N}$ or $\tilde\obsv_\futWind\tilde\reach_\pastWind\tilde Z_{\pastWind,N}$. We will not go in full detail here and remains objective for future research.
\end{rem}

%\pepijn{A comparison in computational time can be found in~\cite[Sec. 9.10]{Cox2018PHD}}.

\section{Simulation Example} \label{sec:SimulationExample} \vspace{-2mm}

In this section, we will demonstrate the performance of the discussed LPV subspace identification schemes on the benchmark example given in~\cite{Verdult2005}. The developed subspace schemes are compared to the $\mbox{PBSID}_{\mbox{opt}}$ method~\cite{Wingerden2009a}.

\vspace{-2mm}
\subsection{Identification setting} \vspace{-2mm}

The benchmark is based on a MIMO LPV-SS model with input dimension $\NU=2$, scheduling dimension $\NP=2$, state dimension $\NX=2$, and output dimension $\NY=2$. To be able to compare the developed approaches to existing LPV subspace methods, we consider the simplified setting with a scheduling independent innovation noise model \vspace{-3mm}
\begin{equation*}
\Kfnc(p_t) = \left[ \begin{array}{cc}
0.32 & 0.16 \\ 
0.64 & 0.24
\end{array}  \right]\!.
\end{equation*}\vskip -6mm \noindent
The innovation noise model is chosen such that the open-loop and closed-loop dynamics are asymptotically input-to-state stable on the domain $p_t\in\mathbb{P}=[-1, 1]^2$ with a quadratic Lyapunov function defined by a constant symmetric matrix.
The noise process $\boldsymbol\xi$ is taken as a white noise with distribution $\boldsymbol\xi_t\sim\mathcal{N}(0,\Xi)$
 where $\Xi$ is diagonal and it is chosen such that the \emph{signal-to-noise ratio} (SNR) %
 \footnote{The noise $w_t$ is a colored noise signal with state-equation $x^w_{t+1}\!=\! \Afnc(p_t)x^w_t\!+\!\Kfnc(p_t)\xi_t$ and output-equation $w_t \!=\! \Cfnc(p_t)x^w_t\!+\!\xi_t$.}
\vspace{-3mm}\begin{equation*}
	\SNR_y^{\ind{i}} = 10 \log_{10} \frac{\sum_{t=1}^N (y^{\ind{i}}_t)^2 }{\sum_{t=1}^N (w^\ind{i}_t)^2},
\end{equation*}\vskip -6mm \noindent
is set for various Monte-Carlo experiments as $\SNR_y^{\ind{i}}=\{\infty,25,10,0\}$\,dB for all $i=1,\ldots,\NY$. The $\ind{i}$ denotes the $i$-th channel, i.e., element of the vector signal, and $\SNR_y^{\ind{i}}$ is the SNR of the output $y^{\ind{i}}$. To evaluate the statistical properties of the subspace schemes, we will carry out two simulation studies with $N=\{10^3,10^4\}$ samples in the identification data set $\Dat$ and in each simulation study $N_{\mathrm{MC}}=100$ Monte Carlo runs are executed. In each run, new realizations of the input and scheduling signals are used.
The simulation output or one-step-ahead predicted output $\hat{y}$ of the estimated model is compared to the measured output and the one-step-ahead predicted output $y$ of the true system (oracle), respectively, by means of the \textit{best fit rate} (BFR)\footnote{Usually the BFR are defined per channel. Eq.~\eqref{eq:BFR} is the average performance criteria over all channels.} \vspace{-2mm}
\begin{align}
\BFR&=\max\!\left\{\hspace{-0.5mm}1\!-\!\frac{\frac{1}{N}\!\sum_{t=1}^N\!\Vert y_t-\hat{y}_t\Vert_2}{\frac{1}{N}\!\sum_{t=1}^N\!\Vert y_t-\bar{y}\Vert_2}, 0\!\right\} \cdot 100\%,\label{eq:BFR}
\end{align}\vskip -5mm \noindent
using a validation data set $\Dval$ as in~\cite{Verdult2005}. %\pepijn{The oracle is the LPV model containing the true underlying parameters and it represents the achievable one-step-ahead prediction of the output signal under the given noise scenario.} 
In \eqref{eq:BFR}, $\bar{y}$ defines the mean of the simulation output or the one-step-ahead predicted output $y$ of the oracle.

For the open-loop setting, the MAX model is estimated using \emph{pseudo linear regression} (PLR) where the update is determined by the enhanced Gauss-Newton method~\cite[Appx. B.3]{Cox2018PHD}. Note that this optimization problem is convex. The PLR is initialized with a FIR model estimate using $\ell_2$-regularized least squares with \emph{generalized cross validation} (GCV) to estimate the optimal regularization parameter~\cite[Sec. 6.1.4]{Golub2013}. For the closed-loop setting, an ARX model is estimated using $\ell_2$-regularized least squares with GCV. The $\mbox{PBSID}_{\mbox{opt}}$~\cite{Wingerden2009a} uses Tikhonov regularization with GCV.

% The future window $\futWind$ and past window $\pastWind$, i.e., the MAX or ARX model orders, are chosen to provide the highest $\BFR$ \TR{on $\Dval$.}
%and, in some cases, a higher model order can be chosen as the estimation variance on the model parameters is lower, e.g., due to the larger dataset of chosen estimation technique or chosen regularization technique.

Next, we will provide a summary of the used design parameters, which are optimized to provide the highest $\BFR$ on $\Dval$. The MAX model orders are $\NB=4$ and $\NC=4$ for $N=10^3$ and $\NB=6$ and $\NC=6$ for $N=10^4$. The ARX model orders are $\NA=6$ and $\NB=6$ for both $N=\{10^3,10^4\}$. The future and past window for the open-loop CCA (Theorem~\ref{lem:LPV-CVA}) are $\futWind=3$ and $\pastWind=4$ for $N=10^3$ and $\futWind=3$ and $\pastWind=5$ for $N=10^4$. For open-loop unified method (Theorem~\ref{lem:UnifiedOL}), the future and past windows are $\futWind=1$ and $\pastWind=4$ for $N=10^3$ and $\futWind=3$ and $\pastWind=4$ for $N=10^4$. The future and past windows for the closed-loop methods (Theorems~\ref{lem:LPV-SSARX} and~\ref{lem:UnifiedCL} and $\mbox{PBSID}_{\mbox{opt}}$) are chosen as $\futWind=3$ and $\pastWind=4$ for both $N=\{10^3,10^4\}$. For the enhanced Gauss-Newton method of the MAX estimation, we enforce the minimal step ascending of $\beta_1 = 10^{-4}$ per iteration (Armijo-Goldstein condition), the initial value for singular value truncation for the search direction is $\gamma = 10^{-8}$, the minimum of the regularization parameter is $\lambda_\mathrm{min} = 10^{-5}$ (Levenberg- Marquardt regularization), the minimum orthogonality requirement of the search direction is $\nu = 0.01$, the minimum step length for backtracking is $\alpha_\mathrm{min} = 0.001$, and the first-order termination condition is $\epsilon = 10^{-6}$ with a maximum of 20 iterations. See~\cite[Algo. 7.1 and Appx. B.3]{Cox2018PHD} for details.

\vspace{-2mm}
\subsection{Analysis of the results} \vspace{-2mm}

\begin{figure*}
\centering
	\begin{subfigure}[b]{0.95\columnwidth}
		\centering
		\input{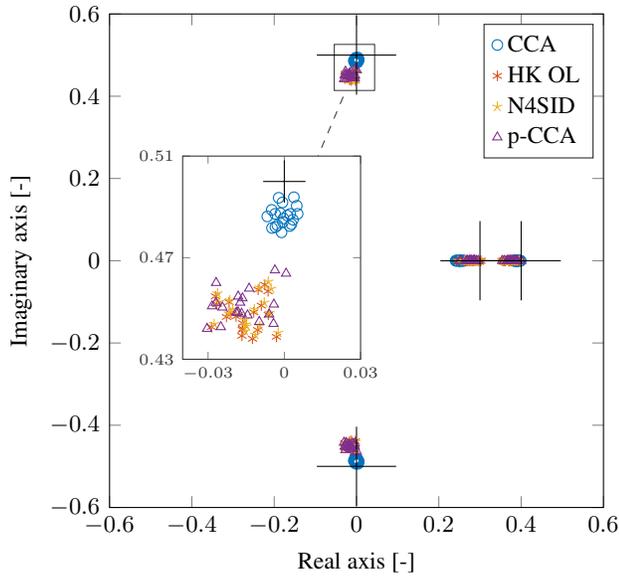}\vspace{-2mm}
		\caption{\vspace{-2mm}Open-loop methodologies.}
	\end{subfigure}\hspace{0.1\columnwidth}
	\begin{subfigure}[b]{0.95\columnwidth}
		\centering
		\input{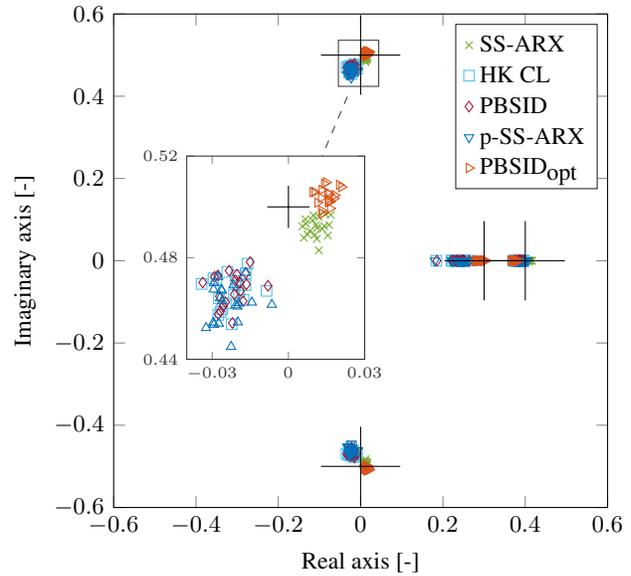}\vspace{-2mm}
		\caption{\vspace{-2mm}Closed-loop methodologies.}
	\end{subfigure}
\caption{The eigenvalues of the estimated $A_0$ and $A_1$ matrices for 20 experiments with $\SNR_y^{\scriptstyle\ind{i}}=10$dB and $N=10^4$ samples in the data set $\Dat$. The figure displays the subspace methodologies using canonical correlation analysis (CCA), Ho-Kalman like projection (HK), Ho-Kalman like projection (HK), N4SID projection, CCA like projection (p-CCA), canonical correlation analysis in closed-loop (SS-ARX) predictor based subspace (PBSID), SS-ARX like projection (p-SS-ARX) and $\mbox{PBSID}_{\mbox{opt}}$ of~\cite{Wingerden2009a}.} \label{fig:eigenval}
\end{figure*}

Table~\ref{tab:results} shows the mean and the standard deviation (between parentheses) of the $\BFR$ on $\Dval$ of the estimation algorithms per Monte Carlo run for various $\SNR_y^{\scriptstyle\ind{i}}=\{\infty,25,10,0\}$dB. The SS-ARX like weighting of Theorem~\ref{lem:UnifiedCL} experiences numerical problems for the data set with $N=10^3$ samples and, therefore, the $\BFR$ is substantially lower. The table shows that the state realization methods based on the maximum-likelihood argument (CCA and SS-ARX) outperform the state realization schemes. Most likely, this difference comes from the fact that the CCA argument obtains a minimum variance estimate of the state given the hypothesized noise. In addition, Figure~\ref{fig:eigenval} also shows that the structural estimation bias of the realization based schemes is bigger than the structural bias of the maximum-likelihood schemes. The structural bias is caused by the fact that the initial condition in Assumption~\ref{ass:initialCond} is not yet small enough. The bias can be further reduced by increasing the past window $\pastWind$; however, this will increase the parameter variance and, therefore, decrease the overall $\BFR$ on the estimate.

Compared with $\mbox{PBSID}_{\mbox{opt}}$ proposed in~\cite{Wingerden2009a}, we can see that direct implementation of the CCA and SS-ARX schemes have comparable performance. However theoretically, $\mbox{PBSID}_{\mbox{opt}}$  should be close to the $\BFR$ performance of the standard implementation of PBSID, but Table~\ref{tab:results} and  Figure~\ref{fig:eigenval} highlight a clear difference which is caused by the kernel trick of $\mbox{PBSID}_{\mbox{opt}}$ that significantly improves numerical accuracy. In addition, the difference in $\BFR$ between some realization techniques is in the order of $10^{-9}$. For example, the case of HK OL, N4SID, and p-CCA for a data set with sample size $N=10^3$. This indicates that the IO estimation step is dominant over the realization step in terms of the $\BFR$ for these particular cases. These observations indicate how important it is to develop a numerically efficient implementation of the developed subspace identification schemes to enhance their performance beyond the theoretical developments of this paper. Therefore, extending the kernel implementation to Theorems~\ref{lem:LPV-CVA}, \ref{lem:UnifiedOL}, \ref{lem:LPV-SSARX} and~\ref{lem:UnifiedCL} is an important objective for future research. Furthermore, while the comparision is provided here with $p$-independent innovation noise models, the developed subspace schemes in this paper are capable to accomplish state estimation with $p$-dependent noise scenarios that are beyond the capabilities of the current state-of-the-art.

%On the other hand, we cannot provide generic statements on the performance of the individual schemes for all identification settings solely based on this single benchmark. Therefore, the performance of these schemes under general conditions is subject for future research.

%\pepijn{Roland can rewrite the above paragraph as you wrote in your comments.}

\begin{table*}[!t] 
\caption{Mean and standard deviation (between parentheses) of the $\BFR$ of the estimation algorithms per Monte Carlo run for different $\SNR_y^{\scriptstyle\ind{i}}=\{\infty,25,10,0\}$dB. The table displays the subspace methodologies using canonical correlation analysis (CCA) in open-loop, the Ho-Kalman like projection (HK) in open-loop (OL) or closed-loop (CL), N4SID projection, CCA like projection (p-CCA), canonical correlation analysis in closed-loop (SS-ARX) predictor based subspace (PBSID), SS-ARX like projection (p-SS-ARX) and $\mbox{PBSID}_{\mbox{opt}}$ of~\cite{Wingerden2009a}.}\label{tab:results} \vspace{0mm}
\centering
	\begin{subfigure}[b]{1\textwidth}
		\caption{$\BFR$ using the simulated output.}
		{\tiny
\setlength\tabcolsep{0.9mm} % between left and right
\begin{tabular}{|l||ll|ll|ll|ll||ll|ll|ll|ll|} \hline
\multirow{2}{*}{$\boldsymbol{\BFR}$ [\%]}&\multicolumn{8}{c||}{$N=10^3$} & \multicolumn{8}{c|}{$N=10^4$} \\ \cline{2-17}
&\multicolumn{2}{c|}{$\infty$dB} & \multicolumn{2}{c|}{$25$dB} & \multicolumn{2}{c|}{$10$dB} & \multicolumn{2}{c||}{$0$dB}  &\multicolumn{2}{c|}{$\infty$dB} & \multicolumn{2}{c|}{$25$dB} & \multicolumn{2}{c|}{$10$dB} & \multicolumn{2}{c|}{$0$dB} \\ \hline
CCA &	$98.45$ 	& ($0.2202$) 	& $97.85$	 & ($0.2568$)	 & $91.10$	 & ($0.8816$)	 & $78.79$	 & ($2.544$)	 & $99.43$	 & ($2.749\cdot 10^{-2}$)	 & $99.05$	 & ($0.0654$)	 & $94.95$	 & ($0.3470$)	 & $87.36$	 & ($1.017$) \\ \hline
HK OL&	 $98.33$ 	& ($0.1940$) 	& $97.90$	 & ($0.2674$)	 & $90.12$	 & ($1.163$)	 & $72.57$	 & ($2.318$)	 & $97.50$	 & ($9.764\cdot 10^{-2}$)	 & $97.40$	 & ($0.1238$)	 & $93.78$	 & ($0.5231$)	 & $78.32$	 & ($1.592$) \\
N4SID &	$98.33$ 	& ($0.1940$) 	& $97.90$	 & ($0.2674$)	 & $90.12$	 & ($1.163$)	 & $72.57$	 & ($2.318$)	 & $97.51$	 & ($9.694\cdot 10^{-2}$)	 & $97.41$	 & ($0.1235$)	 & $93.91$	 & ($0.5031$)	 & $77.32$	 & ($1.716$) \\
p-CCA &	$98.33$ 	& ($0.1940$) 	& $97.90$	 & ($0.2674$)	 & $90.12$	 & ($1.163$)	 & $72.57$	 & ($2.318$)	 & $97.70$	 & ($9.652\cdot 10^{-2}$)	 & $97.50$	 & ($0.1324$)	 & $93.77$	 & ($0.6117$)	 & $74.37$	 & ($2.298$) \\ \hline
SS-ARX &	$95.88$	 & ($0.8259$)	 & $96.02$	 & ($0.7916$)	 & $90.17$	 & ($0.957$)	 & $82.59$	 & ($2.243$)	 & $99.93$	 & ($1.636\cdot 10^{-2}$)	 & $99.36$	 & ($6.611\cdot 10^{-2}$)	 & $95.78$	 & ($0.3136$)	 & $92.71$	 & ($0.5857$) \\ \hline
HK CL &	$94.53$	 & ($1.0340$)	 & $92.39$	 & ($1.397$)	 & $71.75$	 & ($4.340$)	 & $60.72$	 & ($2.645$)	 & $99.85$	 & ($4.567\cdot 10^{-2}$)	 & $96.97$	 & ($0.3260$)	 & $86.15$	 & ($0.9586$)	 & $78.74$	 & ($1.203$)  \\
PBSID &	$94.57$	 & ($1.0205$)	 & $92.45$	 & ($1.382$)	 & $74.55$	 & ($3.162$)	 & $62.16$	 & ($3.255$)	 & $99.85$	 & ($4.567\cdot 10^{-2}$)	 & $97.02$	 & ($0.3223$)	 & $86.27$	 & ($0.9683$)	 & $78.70$	 & ($1.212$)  \\
p-SS-ARX &	$63.02$	 & ($11.79$)	 & $51.18$	 & ($6.672$)	 & $46.00$	 & ($2.728$)	 & $45.38$	 & ($2.664$)	 & $99.86$	 & ($4.510\cdot 10^{-2}$)	 & $96.58$	 & ($0.4401$)	 & $85.64$	 & ($1.005$)	 & $78.50$	 & ($1.228$)\\ \hline 
$\mbox{PBSID}_{\mbox{opt}}$ &	$99.93$	 & ($2.533\cdot 10^{-2}$)	 & $98.98$	 & ($0.1961$)	 & $94.82$	 & ($0.8025$)	 & $86.47$	 & ($1.915$)	 & $99.92$	 & ($1.744\cdot 10^{-2}$)	 & $99.58$	 & ($5.473\cdot 10^{-2}$)	 & $97.53$	 & ($0.2391$)	 & $92.94$	 & ($0.5976$) \\ \hline 
\end{tabular}
}
	\end{subfigure}
	\begin{subfigure}[b]{1\textwidth}
		\caption{$\BFR$ using the predicted output.}
		{\tiny
\setlength\tabcolsep{0.9mm} % between left and right
\begin{tabular}{|l||ll|ll|ll|ll||ll|ll|ll|ll|} \hline
\multirow{2}{*}{$\boldsymbol{\BFR}$ [\%]}&\multicolumn{8}{c||}{$N=10^3$} & \multicolumn{8}{c|}{$N=10^4$} \\ \cline{2-17}
&\multicolumn{2}{c|}{$\infty$dB}  & \multicolumn{2}{c|}{$25$dB} & \multicolumn{2}{c|}{$10$dB} & \multicolumn{2}{c||}{$0$dB}  &\multicolumn{2}{c|}{$\infty$dB} & \multicolumn{2}{c|}{$25$dB} & \multicolumn{2}{c|}{$10$dB} & \multicolumn{2}{c|}{$0$dB} \\ \hline
CCA &	 $98.92$ & ($0.1708$)	 & $96.21$	 & ($0.1470$)	 & $85.20$	 & ($0.8021$)	 & $65.47$	 & ($1.843$)	 & $99.50$	 & ($2.504\cdot 10^{-2}$)	 & $97.00$	 & ($3.581\cdot 10^{-2}$)	 & $87.00$	 & ($0.2044$)	 & $67.88$	 & ($0.4496$) \\ \hline
HK OL &	$98.90$ & ($0.1557$)	 & $96.27$	 & ($0.1596$)	 & $84.17$	 & ($0.8873$)	 & $58.84$	 & ($1.592$)	 & $97.63$	 & ($0.1072$)	 & $95.65$	 & ($8.794\cdot 10^{-2}$)	 & $84.15$	 & ($0.6078$)	 & $57.24$	 & ($0.8006$) \\
N4SID  &	$98.90$ & ($0.1557)$	 & $96.27$	 & ($0.1596$)	 & $84.17$	 & ($0.8873$)	 & $58.84$	 & ($1.592$)	 & $97.63$	 & ($0.1068$)	 & $95.65$	 & ($8.748\cdot 10^{-2}$)	 & $84.13$	 & ($0.6025$)	 & $57.37$	 & ($0.9458$) \\
p-CCA &	$98.90$ & ($0.1557$)	 & $96.27$	 & ($0.1596$)	 & $84.17$	 & ($0.8873$)	 & $58.84$	 & ($1.592$)	 & $97.72$	 & ($0.1085$)	 & $95.68$	 & ($9.926\cdot 10^{-2}$)	 & $84.79$	 & ($0.6815$)	 & $57.60$	 & ($1.215$) \\ \hline
SS-ARX &	$97.24$	 & ($0.5553$)	 & $95.31$	 & ($0.5791$)	 & $84.30$	 & ($1.064$)	 & $72.19$	 & ($1.657$)	 & $99.94$	 & ($1.323\cdot 10^{-2}$)	 & $97.02$	 & ($0.0760$)	 & $88.63$	 & ($0.2212$)	 & $77.42$	 & ($0.5231$) \\ \hline
HK CL&	$95.86$	 & ($0.7027$)	 & $93.17$	 & ($0.8058$)	 & $72.53$	 & ($2.883$)	 & $51.40$	 & ($2.905$)	 & $99.88$	 & ($3.682\cdot 10^{-2}$)	 & $96.75$	 & ($0.2393$)	 & $85.93$	 & ($0.6049$)	 & $72.27$	 & ($0.6701$) \\
PBSID &	$95.88$	 & ($0.7007$)	 & $93.20$	 & ($0.798$)	 & $74.50$	 & ($2.093$)	 & $53.48$	 & ($3.061$)	 & $99.88$	 & ($3.682\cdot 10^{-2}$)	 & $96.77$	 & ($0.2351$)	 & $85.98$	 & ($0.6128$)	 & $72.25$	 & ($0.6748$) \\
p-SS-ARX &	$68.09$	 & ($12.87$)	 & $54.42$	 & ($8.925$)	 & $44.78$	 & ($3.105$)	 & $29.11$	 & ($1.757$)	 & $99.88$	 & ($3.627\cdot 10^{-2}$)	 & $95.58$	 & ($0.3735$)	 & $84.79$	 & ($0.7385$)	 & $71.56$	 & ($0.7321$) \\ \hline
$\mbox{PBSID}_{\mbox{opt}}$ &	$99.95$	 & ($1.710\cdot 10^{-2}$)	 & $97.23$	 & ($0.1720$)	 & $87.95$	 & ($0.5796$)	 & $74.09$	 & ($1.624$)	 & $99.94$	 & ($1.397\cdot 10^{-2}$)	 & $97.12$	 & ($0.0689$)	 & $89.13$	 & ($0.1646$)	 & $76.34$	 & ($0.3061$)\\ \hline
\end{tabular}
}

	\end{subfigure}
\end{table*}

\section{Conclusion} \vspace{-2mm}

In this paper, we have presented a unified framework to formualte extension of supace identification methods for LPV identifcation by systematically developping an LPV subspace identification theory. Based on the derived open-loop, closed-loop, and predictor-based data-equations, several methods have been proposed to estimate LPV-SS models in one unified framework based on a maximum-likelihood or realization based arguments. Hence, we have shown how to extend LTI CVA, SS-ARX, PBSID, and N4SID to the LPV setting. %Furthermore, the number of to-be-estimated parameters and the dimensions of the matrices is reduced for the subspace identification problem and, hence, the schemes presented have a lower computational demand. 
The effectiveness of the presented subspace identification methods is demonstrated in a Monte Carlo study by identifying a MIMO LPV benchmark system. An important future direction of research is to imporve nummerical efficency and reduce computational load of the developped methods.

\begin{ack}                               % Place acknowledgements here.
\vspace{-2mm}
{\small We would like to thank the authors of~\cite{Wingerden2009a} for providing their code to make the simulation study possible.}
\end{ack}

\bibliographystyle{ieeetr}        % Include this if you use bibtex 
\bibliography{library.bib}           % and a bib file to produce the 
                                 % bibliography (preferred). The
                                 % correct style is generated by
                                 % Elsevier at the time of printing.

\appendix

\section{Proof of Lemma~\ref{lem:LPV-CVA}} \label{app-sec:ProofCCA} \vspace{-2mm}

The idea of the proof is based on the reasoning used in~\cite{Gicans2009} for the LTI case, however, we will also fix some of the inconsistencies found in~\cite{Gicans2009}. Representation~\eqref{eq:DataEqAffOLTIObsvWithCorrFut} reads as: \vspace{-3mm}
\begin{equation} \label{eq:DataEqAffOLTIObsvWithCorrFut1}
\futDatcCor{y}{\futWind}{t} = \underbrace{\obsv^0_\futWind\check\reach_\pastWind}_{\hank^0_{\futWind,\pastWind}} \check M_{t,\pastWind}\pastDatc{z}{t}{\pastWind} +  \futDat{\xi}{\futWind}{t} = \obsv^0_\futWind x_t +\futDat{\xi}{\futWind}{t}.
\end{equation}\vskip -8mm \noindent
%
%\begin{equation} \label{eq:DataEqAffOLTIObsvWithCorrFut1}
%y_t = \underbrace{\obsv_\futWind\reach_\pastWind}_{\hank_{\futWind,\pastWind}} z_t +  \xi_t =\obsv_\futWind x_t +  \xi_t, \mbox{ with }z_t=M_{t}\futDatc{z}{t}{},
%\end{equation}
%
Note that $\futDat{\xi}{\futWind}{t}$ is a sample path realization of a Gaussian white noise with variance $\Sigma^2_\xi=\eye_\futWind\otimes\Xi^2$. The model~\eqref{eq:DataEqAffOLTIObsvWithCorrFut1} is dependent on the unknown sub-Markov parameters in $\hank^0_{\futWind,\pastWind}$, which are parameterized as $\hank$ and the unknown noise variance $\Sigma^2_\xi$ with symmetric parameterization $\Sigma^2$. %\pepijn{Note that $\hank$ and $\Sigma^2_\xi$ are matrices.} 
Due to the Markov property of representation~\eqref{eq:DataEqAffOLTIObsvWithCorrFut1} and by employing Bayes' rule, the maximum likelihood estimate can be obtained by maximizing the likelihood or, equivalently, by minimizing the log-likelihood based on the data set $\Dat$ \vspace{-4mm}
\begin{multline} \label{eq:minLogLikeLS}
\min_{\substack{\hank\in\Theta\\ \Sigma^2\in\sS}} -\log L(\hank,\Sigma^2) = \min_{\substack{\hank\in\Theta\\ \Sigma^2\in\sS}}  \frac{1}{2} \futWind\,\NY N \log(2\pi) \\
+\frac{1}{2} N  \log\left( \det\left( \Sigma^2 \right)  \right) + \frac{1}{2} \sum_{t=1}^{N} \varepsilon^\top_{t|\hank} \Sigma^{-2} \varepsilon_{t|\hank},
\end{multline}\vskip -6mm \noindent
where the one-step-ahead prediction-error is $\varepsilon_{t|\hank} = \futDatcCor{y}{\futWind}{t} - \hank\check M_{t,\pastWind}\pastDatc{z}{t}{\pastWind}$ and the future window is $\futWind$. As the signals are assumed to be persistently exciting, i.e., $\check Z_{\pastWind,N}(\check Z_{\pastWind,N})^\top\succ0$, the well-known unique stationary point of~\eqref{eq:minLogLikeLS} is obtained at \cite[Lem. 3.3]{Gibson2005} %
\footnote{
The estimate $\hat\Sigma_\xi^2 = \frac{1}{N} \varepsilon_{t|\hat\hank} \varepsilon^\top_{t|\hat\hank} $ is simplified as \vspace{-4mm} %by recognizing the following
\begin{multline*}
\frac{1}{N}\check Z_{\pastWind,N} (\check Y_{\futWind,N}^{(\mathrm{c})}-\hat\hank_{\futWind,\pastWind} \check Z_{\pastWind,N})^\top = \frac{1}{N}\check Z_{\pastWind,N} \Bigl(\check Y_{\futWind,N}^{(\mathrm{c})} \\
-  \frac{1}{N}\check Y_{\futWind,N}^{(\mathrm{c})} \check Z_{\pastWind,N}^\top\! \Bigl( \frac{1}{N} \check Z_{\pastWind,N} \check Z_{\pastWind,N}^\top\Bigr)^{-1} \check Z_{\pastWind,N}\Bigr)^\top =0. %\\
%&= \frac{1}{N}Z_N Y^\top_N - \frac{1}{N}Z_NZ^\top_{N}\Bigl(\frac{1}{N}  Z_N Z^\top_{N}\Bigr)^{-1} \frac{1}{N} Z_NY^\top_N =0.
\end{multline*}\vskip -6mm \noindent
} %
\begin{subequations}\label{eq:estLSHankSigma}
\vspace{-3mm}\begin{align} 
\hat\hank_{\futWind,\pastWind}^0&\!=\!\! \frac{1}{N} \check Y_{\futWind,N}^{(\mathrm{c})} \check Z_{\pastWind,N}^\top(\check Z_{\pastWind,N}\check Z_{\pastWind,N}^\top)^{-1}, \\
 \hat\Sigma_\xi^2 &\!=\!\!  \frac{1}{N} \check Y_{\futWind,N}^{(\mathrm{c})} (\check Y_{\futWind,N}^{(\mathrm{c})})^{\!\top} \!\!\!-\! \frac{1}{N} \hat\hank_{\futWind,\pastWind}^0 \check Z_{\pastWind,N}\check Z_{\pastWind,N}^\top(\hat\hank_{\futWind,\pastWind}^0)^{\!\top}\!\!\!.
\end{align}\vskip -6mm \noindent
\end{subequations}
The solution~\eqref{eq:estLSHankSigma} is a consistent estimate which is efficient in terms of the parameter variance. The interest is not in an estimate of $\hank_{\futWind,\pastWind}^0$, but to attain a realization of the state, $\obsv^0_\futWind$, and $\check\reach_\pastWind$ separately, that together will maximize the likelihood~\eqref{eq:minLogLikeLS}. To start, assume that $\check\reach_\pastWind \check Z_{\pastWind,N}$  is known a-priori in~\eqref{eq:DataEqAffOLTIObsvWithCorrFut1}. Then, similar to~\eqref{eq:minLogLikeLS}, the solution to the linear least-squares problem is \vspace{-3mm}
\begin{subequations}\label{eq:CVALSpart2}
\begin{align}
\hat{\obsv}_\futWind^0 &\!=\! \frac{1}{N}  \check Y_{\futWind,N}^{(\mathrm{c})} \check Z_{\pastWind,N}^{\!\top}\check\reach^\top_\pastWind \!\!\left( \frac{1}{N}  \check\reach_\pastWind \check Z_{\pastWind,N} \check Z_{\pastWind,N}^\top \check\reach_\pastWind^\top\right)^{\!-1}\!\!\!\!, \label{eq:estObsv1} \\
\hat{\Sigma}^2_\xi &\!=\! \frac{1}{N} \check Y_{\futWind,N}^{(\mathrm{c})}(\check Y_{\futWind,N}^{(\mathrm{c})})^{\!\top} - \frac{1}{N^2} \check Y_{\futWind,N}^{(\mathrm{c})} \check Z_{\pastWind,N}^\top \check\reach^\top_\pastWind \nonumber\\
&\hspace{0.2cm}\times\left( \frac{1}{N}  \check\reach_\pastWind \check Z_{\pastWind,N} \check Z_{\pastWind,N}^\top \check\reach_\pastWind^\top\right)^{\!-1}\!\!\check\reach_\pastWind \check Z_{\pastWind,N} (\check Y_{\futWind,N}^{(\mathrm{c})})^{\!\top}. \label{eq:estNoise}
\end{align}\vskip -7mm \noindent
\end{subequations}
The log-likelihood function associated with~\eqref{eq:CVALSpart2}, given $\check\reach_\pastWind$, $\check Z_{\pastWind,N}$ and $\check Y_{\futWind,N}^{(\mathrm{c})}$, is \vspace{-3mm}
\begin{multline} \label{eq:LogLikeLS1}
-\log L(\hat{\obsv}_\futWind^0\check\reach_\pastWind,\hat\Sigma^{2}_\xi) = \frac{1}{2} \futWind\,\NY N \left( \log(2\pi) + 1\right) \\ 
+\frac{N}{2}  \log\biggl(\det\biggl( \frac{1}{N} \check Y_{\futWind,N}^{(\mathrm{c})}(\check Y_{\futWind,N}^{(\mathrm{c})})^\top - \frac{1}{N^2} \check Y_{\futWind,N}^{(\mathrm{c})} \check Z_{\pastWind,N}^\top \check\reach^\top_\pastWind\! \\ 
\times \left( \frac{1}{N}  \check\reach_\pastWind \check Z_{\pastWind,N}\check Z_{\pastWind,N}^\top\reach_\pastWind^\top\right)^{-1}\!\!\check\reach_\pastWind \check Z_{\pastWind,N} (\check Y_{\futWind,N}^{(\mathrm{c})})^\top \biggr)  \biggr).
\end{multline}\vskip -7mm \noindent
Note that the last product in~\eqref{eq:minLogLikeLS} can be simplified to \vspace{-3mm}
\begin{equation*}
 \frac{1}{2} \sum_{t=1}^{N}\! \varepsilon^\top_{t|\hat{\obsv}_\futWind^0\check\reach_\pastWind} \hat\Sigma^{-2}_\xi \varepsilon_{t|\hat{\obsv}_\futWind^0\check\reach_\pastWind} 
\!=\! \frac{1}{2}\Tr\left\{ N \hat\Sigma^{2}_\xi  \hat\Sigma^{-2}_\xi \right\} \!=\! \frac{1}{2} \futWind\,\NY N.
\end{equation*}\vskip -7mm \noindent
Next, the focus will be on obtaining the $\pastWind$-step extended reachability matrix $\check\reach_\pastWind$ that minimizes the log-likelihood~\eqref{eq:LogLikeLS1}. Similar to~\cite{Larimore2005}, take the following constrained  SVD \vspace{-4mm}
\begin{equation}\label{eq:GSVD}
\begin{aligned} 
&&\tilde S &= \tilde U^\top \frac{1}{N} \check Y_{\futWind,N}^{(\mathrm{c})}\check Z_{\pastWind,N}^\top \tilde V, \\
\mbox{s.t.}\hspace{0.2cm} && \eye_{\NY\futWind} &= \tilde U^\top \frac{1}{N} \check Y_{\futWind,N}^{(\mathrm{c})}(\check Y_{\futWind,N}^{(\mathrm{c})})^\top \tilde U, \\
&&\eye_\NZ &= \tilde V^\top \frac{1}{N} \check Z_{\pastWind,N} \check Z_{\pastWind,N}^\top \tilde V,
\end{aligned}
\end{equation}\vskip -6mm \noindent
where $\tilde U\in\mathbb{R}^{\NY\futWind\times\NY\futWind}$,  $\tilde S\in\mathbb{R}^{\NY\futWind\times\NZ}$, and $\tilde V\in\mathbb{R}^{\NZ\times\NZ}$, and $\NZ=(\NU+\NY)\sum_{i=1}^\pastWind(1+\NPSI)^i$. This constrained SVD is unique in case $\check Y_{\futWind,N}^{(\mathrm{c})}(\check Y_{\futWind,N}^{(\mathrm{c})})^\top\succ0$ and $\check Z_{\pastWind,N} \check Z_{\pastWind,N}^\top\succ0$, which comes naturally under the persistency of excitation conditions. The matrix $\tilde S$ is a diagonal matrix with ordered singular values $\tilde \sing^2_1\geq\ldots\geq\tilde \sing^2_n\geq0$ and $n=\min(\NY\futWind,\NZ)$. In the following discussion, we assume that $\NY\futWind\leq\NZ$ to simplify notation. Note that~\eqref{eq:GSVD} is equivalent to \vspace{-4mm}
\begin{equation} \label{eq:SVDcov}
\begin{aligned}
\frac{1}{N} \check Y_{\futWind,N}^{(\mathrm{c})}(\check Y_{\futWind,N}^{(\mathrm{c})})^\top &\!=\! (\tilde U \tilde U^\top)^{-1}, \hspace{0.2cm}\frac{1}{N} \check Z_{\pastWind,N} \check Z_{\pastWind,N}^\top &&\hspace{-4mm}\!=\! (\tilde V \tilde V^\top)^{-1}, \\
\frac{1}{N}\check Y_{\futWind,N}^{(\mathrm{c})}\check Z_{\pastWind,N}^\top &\!=\! (\tilde U \tilde U^\top)^{-1} \tilde U \tilde S \tilde V^\top(\tilde V \tilde V^\top)^{-1} &&\hspace{-4mm}\!=\! \tilde U^\dagger \tilde S (\tilde V^\dagger)^{\!\top}\!,
\end{aligned}
\end{equation}
by applying the left pseudo-inverses. Substituting the relations~\eqref{eq:SVDcov} into the second line of~\eqref{eq:LogLikeLS1} leads to \vspace{-3mm}
\begin{multline} \label{eq:CCAinterMedRes1}
%\frac{N}{2}  \log\Big( \det\Big( (\tilde U \tilde U^\top)^{-1}  - \tilde U^\dagger \tilde S (\tilde V^\dagger)^{\!\top} \check\reach^\top_\pastWind \\
%\times \left(\! \check\reach_\pastWind (\tilde V \tilde V^\top)^{-1} \check\reach_\pastWind^\top\!\right)^{-1} \check\reach_\pastWind \tilde V^\dagger \tilde S^\top (\tilde U^\dagger)^{\!\top}  \Big)  \Big) \\
% =
\frac{N}{2}  \log\Big( \det\bigr( \tilde U^\dagger\bigl) \det\Big( \! \eye_{\NY\futWind}  - \tilde S  (\tilde V^\dagger)^{\!\top} \check\reach^{\!\top}_\pastWind   \\
\!\times\! \left( \check\reach_\pastWind (\tilde V \tilde V^\top)^{-1} \check\reach_\pastWind^\top\right)^{\!-1}\! \check\reach_\pastWind \tilde V^\dagger \tilde S^\top \Big)  \det\bigl( (\tilde U^\dagger)^{\!\top} \bigr)  \Big).
\end{multline}\vskip -7mm \noindent
In~\eqref{eq:CCAinterMedRes1}, the product property of the determinant is applied: $\det(AB)=\det(A)\cdot\allowbreak\det(B)$. It is important to see that the constrained SVD~\eqref{eq:GSVD} only decomposes the signal relations based on the (co)variances and the decomposition does not change the minimization of~\eqref{eq:LogLikeLS1}. To simplify the notation, define \vspace{-1mm}
\begin{equation} \label{eq:detrmQ}
Q \triangleq  (\tilde V^\dagger)^\top \check\reach_\pastWind^\top\in\mathbb{R}^{\NZ\times\NX},
\end{equation}\vskip -5mm \noindent
which represents an injective mapping of $\check\reach_\pastWind$ to $Q$ as $\tilde V$ is full rank. Applying this transformation,~\eqref{eq:CCAinterMedRes1} reads as \vspace{-3mm}
\begin{multline} \label{eq:CCAinterMedRes2}
-\frac{1}{2} N  \log\Big( \det\Big( (\tilde U \tilde U^\top)^{-1}  \Big) \\ \times
 \det\Big(  \eye_{\NY\futWind}-\tilde S Q (Q^\top Q)^{-1} Q^\top\tilde S^\top \Big)\Big).
\end{multline}\vskip -7mm \noindent
Note that minimization of~\eqref{eq:LogLikeLS1} is equivalent to minimizing~\eqref{eq:CCAinterMedRes2} with a change of variables. Furthermore, the inverted expression $Q^\top Q$ in~\eqref{eq:CCAinterMedRes2} can be written as \vspace{-3mm}
\begin{multline} \label{eq:stateIdent}
Q^\top Q = \check\reach_\pastWind (\tilde V \tilde V^\top)^{-1} \check\reach_\pastWind^\top \\
= \frac{1}{N} \check\reach_\pastWind \check Z_{\pastWind,N}\check Z_{\pastWind,N}^\top  \check\reach_\pastWind^\top = \frac{1}{N}X_NX_N^\top,
\end{multline}\vskip -7mm \noindent
which is the sample variance of the to-be-chosen state variable. The realized state is an isomorphic representation with respect to the original SS form of the underlying data-generating system, see Section~\ref{sec:ParamSID}. Hence, its sample variance $\frac{1}{N}X_NX_N^\top$ can be chosen to be any arbitrary positive definite matrix. In CCA, the variance is chosen to be identity, hence, all states have equal magnitude and are maximally uncorrelated.  As such, minimizing the log-likelihood~\eqref{eq:CCAinterMedRes2} renders to \vspace{-4mm}
\begin{multline} \label{eq:CCAinterMedRes3}
\min_{Q^\top Q = \eye_\NX} \!\!\log\!\Big(\!\! \det\!\Big(\!  \eye_{\NY\futWind}-\tilde SQ Q^\top\tilde S^\top \!\Big)\!\Big) \\
=\min_{Q^\top Q = \eye_\NX}\!\! \log\!\Big(\!\! \det\!\Big(\!  \eye_{\NY\futWind}- Q^\top\tilde S^\top\tilde SQ \!\Big)\!\Big),
\end{multline}\vskip -7mm \noindent
by applying Sylvester's determinant identity. To find an expression for minimizing the log-likelihood, note that the determinant of a matrix is the product of its eigenvalues. From Poincar\'{e} separation theorem (Lemma~\ref{lem:Poincare}) it follows that the choice $Q^\top Q = \eye_\NX$ will not lead to a single solution of~\eqref{eq:CCAinterMedRes3}. Using Lemma~\ref{lem:Poincare}, let us investigate the individual descending sorted eigenvalues \vspace{-4mm}
\begin{multline} \label{eq:equalityEigenValues}
\eig_{\NY\futWind-i+1}\bigl\{ \eye_\NX - Q^\top\tilde S^\top \tilde S Q \bigr\} = 1-\eig_i\bigl\{Q^\top\tilde S^\top \tilde S Q \bigr\} \\
 \geq 1-\eig_i\bigl\{\tilde S^\top \tilde S \bigr\} = 1- \tilde \sing^2_i,
\end{multline}\vskip -6mm \noindent
for $i=1,\ldots,\NX$. Hence, minimization of the marginal likelihood~\eqref{eq:CCAinterMedRes3} has a lower-bound based on the product of the singular values in $\tilde S$. The lower-bound is clearly obtained if $Q = [ \begin{array}{cc} \eye_\NX & 0 \end{array}]^\top$, which also satisfies $Q^\top Q=\eye_\NX$. There might be other solutions to the minimization problem of~\eqref{eq:CCAinterMedRes3}, however, we take the solution equal to the CVA solution of~\cite{Larimore1983}. Hence, the latter choice of $Q$ maximizes the marginal likelihood function~\eqref{eq:LogLikeLS1}. An estimate of the reachability matrix  $\hat\reach_\pastWind$ is obtained by reformulating~\eqref{eq:detrmQ} as \vspace{-3mm}
\begin{equation} \label{eq:estReachCCA}
\hat\reach_\pastWind = Q^\top \tilde V^\top=[ \begin{array}{cc} \eye_\NX & 0 \end{array}]\tilde V^\top,
\end{equation}\vskip -7mm \noindent
which results in selecting the first $\NX$ columns of $\tilde V$. Then, the estimates of the observability \eqref{eq:estObsv1} and noise covariance \eqref{eq:estNoise} are equivalent to \vspace{-3mm}
\begin{subequations}
\begin{align}
\hat\obsv_\futWind &= \tilde U^\dagger \tilde S (\tilde V^\dagger)^{\!\top} \hat\reach_\pastWind^\top =  \tilde U^\dagger \tilde S Q, \label{eq:estObsvCCA} \\
\hat\Sigma_\xi^2 &= (\tilde U \tilde U^\top)^{-1} -  \tilde U^\dagger \tilde S (\tilde V^\dagger)^{\!\top}\hat\reach^\top_\pastWind \hat\reach_\pastWind \tilde V^\dagger \tilde S^\top (\tilde U^\dagger)^{\!\top} \nonumber \\
&= \tilde U^\dagger (\eye_{\NY\futWind} - \tilde S Q Q^\top \tilde S^\top) (\tilde U^\dagger)^{\!\top}, \label{eq:estNoiseCCA}
\end{align}\vskip -7mm \noindent
\end{subequations}
where $(\tilde V^\dagger)^{\!\top}\tilde V= \frac{1}{N}\check\reach_\pastWind \check Z_{\pastWind,N}\check Z_{\pastWind,N}^\top \check\reach^\top_\pastWind=\eye_\NX$ due to~\eqref{eq:GSVD} and~\eqref{eq:stateIdent}. Note that the multiplication $\tilde S Q$ selects only the first $\NX$ singular values of $\tilde S$. Combining the estimates~\eqref{eq:estReachCCA} and~\eqref{eq:estObsvCCA} results in an estimate of $\hank^0_{\futWind,\pastWind}$: \vspace{-3mm}
\begin{equation} \label{eq:hankEstCCA1}
\hat\hank^0_{\futWind,\pastWind}= \tilde U^\dagger \tilde S Q Q^\top \tilde V^\top.
\end{equation}\vskip -7mm \noindent
The log-likelihood function corresponding to the estimates~\eqref{eq:estReachCCA}-\eqref{eq:hankEstCCA1} is \vspace{-3mm}
\begin{multline*}
-\log L(\hat\hank_{\futWind,\pastWind}^0,\hat\Sigma^{2}_\xi) = \frac{\futWind\,\NY N}{2} \left( \log(2\pi) + 1\right) \\
+ \frac{N}{2}  \log\!\left(\! \det\!\left( \frac{1}{N} \check Y_{\futWind,N}^{(\mathrm{c})} (\check Y_{\futWind,N}^{(\mathrm{c})})^\top\right) \! \prod_{i=1}^\NX (1-\tilde \sing^2_i)  \right) = \\ 
- \frac{N}{2} \log\bigl(\det\bigl(\tilde U\bigr)^2\bigr) + \frac{\futWind\,\NY N}{2} \left( \log(2\pi) \!+\! 1\right) + \frac{N}{2} \!\sum_{i=1}^\NX  \log(1-\tilde \sing^2_i ).
\end{multline*}\vskip -7mm \noindent
The last remaining step is to show that the estimates of the Hankel matrix $\hat\hank_{\futWind,\pastWind}^0$~\eqref{eq:hankEstCCA1} and the noise $\hat\Sigma_\xi^2$~\eqref{eq:estNoiseCCA} are equivalent to the estimates in~\eqref{eq:estLSHankSigma} that characterize the minimum of~\eqref{eq:minLogLikeLS}. Substitute~\eqref{eq:SVDcov} in~\eqref{eq:estLSHankSigma} \vspace{-3mm}
\begin{subequations} \label{eq:estLSHankSigma1}
\begin{align}
\hat\hank_{\futWind,\pastWind}^0&\!=\! \frac{1}{N} \check Y_{\futWind,N}^{(\mathrm{c})} \check Z_{\pastWind,N}^\top(\frac{1}{N}\check Z_{\pastWind,N}\check Z_{\pastWind,N}^\top)^{-1} \!=  \tilde U^\dagger \tilde S \tilde V^\top\!\!, \\
\hat\Sigma_\xi^2 &\!=\!   \frac{1}{N} \check Y_{\futWind,N}^{(\mathrm{c})}(\check Y_{\futWind,N}^{(\mathrm{c})})^\top \!-\! \frac{1}{N} \hat\hank_{\futWind,\pastWind}^0 \check Z_{\pastWind,N}\check Z_{\pastWind,N}^\top \hat\hank_{\futWind,\pastWind}^\top \nonumber \\
&= \tilde U^\dagger \bigl(\eye_{\NY\futWind} \!-\! \tilde S \tilde S^{\!\top}\bigr) (\tilde U^\dagger)^{\!\top}.
\end{align}\vskip -6mm \noindent
\end{subequations}
The estimates~\eqref{eq:estNoiseCCA} and~\eqref{eq:hankEstCCA1} are identical to~\eqref{eq:estLSHankSigma1} when the all singular values are selected, due to $\tilde S Q$. In case the number of  data points goes to infinity, i.e., $N\rightarrow\infty$, then $\tilde S$ will contain exactly $\NX$ nonzero singular values. In the finite data case, the number of states in the realization is selected based on $Q$. In conclusion, the SVD~\eqref{lem-eq:loglikelihoodCVA} maximizes the marginal-likelihood function of the linear estimation problem~\eqref{eq:DataEqAffOLTIObsvWithCorrFut} w.r.t.\ the unknowns $\obsv^0_\futWind$, $\check\reach_\pastWind$ and the covariance $\Xi^2$ with state-sequence~\eqref{lem-eq:CCAstate} and log-likelihood function~\eqref{lem-eq:loglikelihoodCVA}.

Note that in early literature on CVA SID \cite{Larimore1983,Larimore1990b}, the constrained SVD ~\eqref{eq:GSVD} was performed with arbitrary positive-definite weight $\Lambda\in\sS$ such that $\eye = \tilde V^\top \Lambda \tilde V$, which is called the CVA method. The CCA and CVA method coincides with the weighting choice in~\eqref{eq:GSVD}. The CVA method with $\Lambda\neq \frac{1}{N} \check Z_{\pastWind,N} (\check Z_{\pastWind,N})^\top$ leads to a minimal prediction-error solution \cite[Eq. (10)]{Larimore1990b}, but will not lead to a maximum-likelihood estimate.
 \hfill $\blacksquare$

\begin{lem}[Poincar\'{e} separation theorem] \label{lem:Poincare}
Let $A\in\sS^n$ and $B\in\mathbb{R}^{n\times r}$ be matrices such that $B^\top B=\eye_r$. Let $\eig_i\{\centerdot\}$ represent the eigenvalues of a matrix sorted in descending order. Then, \vspace{-3mm}
\begin{equation*}
\eig_i\left\{ B^{\!\top}\! AB \right\} \leq \eig_i\left\{A\right\}, \hspace{1cm} i=1,\ldots,r.
\end{equation*} \vskip -10mm
\hfill $\square$
\end{lem}\vspace{-5mm}
\begin{pf}
See~\cite[p. 64]{Rao1973}.\hfill $\blacksquare$
\end{pf}

\end{document}